\documentclass[11pt, a4paper, oneside, reqno]{amsart}
\usepackage{comment}
\usepackage[usenames, dvipsnames]{color}
\definecolor{darkblue}{rgb}{0.0, 0.0, 0.45}
\definecolor{lightblue}{RGB}{240,248,255}
\definecolor{lightblue2}{rgb}{0.68, 0.85, 0.9}
\definecolor{lightcyan}{rgb}{0.88, 1.0, 1.0}
\definecolor{palepink}{rgb}{0.98, 0.85, 0.87}

\usepackage[colorlinks	= true,
raiselinks	= true,
linkcolor	= darkblue, 
citecolor	= Mahogany,
urlcolor	= ForestGreen,
pdfauthor	= {Peyman Mohajerin Esfahani},
pdftitle	= {},
pdfkeywords	= {},
pdfsubject	= {},
plainpages	= false]{hyperref}

\usepackage{dsfont,amssymb,amsmath,enumitem} 
\usepackage{amsfonts,dsfont,mathtools, mathrsfs,amsthm} 
\usepackage[amssymb, thickqspace]{SIunits}
\usepackage{fancyhdr,mdframed,nicefrac}

\usepackage{epsfig}
\usepackage{graphicx}
\usepackage{float}
\usepackage{caption}
\usepackage{subcaption}

\usepackage{courier}

\usepackage{multirow}
\usepackage{bigstrut}

\allowdisplaybreaks
\date{\today}
\makeatletter
\def\@settitle{\begin{center}%
		\baselineskip14\p@\relax
		\normalfont\LARGE\scshape\bfseries
		\@title
	\end{center}%
}

\def\@setauthors{%
  \begingroup
  \def\thanks{\protect\thanks@warning}%
  \trivlist
  \centering\footnotesize \@topsep30\p@\relax
  \advance\@topsep by -\baselineskip
  \item\relax
  \author@andify\authors
  \def\\{\protect\linebreak}%
  \authors%
  \ifx\@empty\contribs
  \else
    ,\penalty-3 \space \@setcontribs
    \@closetoccontribs
  \fi
  \endtrivlist
  \endgroup
}

\makeatother
\makeatletter

\def\subsection{\@startsection{subsection}{2}%
	\z@{.5\linespacing\@plus.7\linespacing}{.5\linespacing}%
	{\normalfont\large\bfseries}}

\def\subsubsection{\@startsection{subsubsection}{3}%
	\z@{.5\linespacing\@plus.7\linespacing}{.5\linespacing}%
	{\normalfont\itshape}}

\usepackage{multirow}
\usepackage{bigstrut}
\usepackage{wrapfig}
\usepackage{caption}

\renewcommand{\ge}{\geqslant}

\DeclareSymbolFont{symbolsC}{U}{pxsyc}{m}{n}

\DeclareMathOperator*{\argmin}{argmin}

\usepackage{bm}
\usepackage{amsthm}

\usepackage{amssymb}
\usepackage{latexsym}
\usepackage{mathrsfs}  
\usepackage{bbm}
\usepackage{amsthm}

\usepackage{url}
\usepackage{color}

\usepackage{amsmath,bm}
\usepackage{subcaption}
\usepackage[makeroom]{cancel}

\usepackage{float}
\usepackage{lipsum}
\usepackage{dsfont}
\usepackage{rotating}
\usepackage{mathtools}
\usepackage{placeins}
\usepackage{relsize}

\title[Coupling kinetic and continuum using data-driven maximum entropy distribution]{
Coupling kinetic and continuum using data-driven maximum entropy distribution
}
 \author{Mohsen Sadr, Qian Wang, and M. Hossein Gorji}
  \thanks{Corresponding author: Mohsen Sadr}
 \thanks {Email: mohsen.sadr@epfl.ch}
\thanks{Mohsen Sadr: Applied and Computational Mathematics, RWTH Aachen University, Schinkestrasse 2, D-52062 Aachen, Germany and Ecole Polytechnique F{\'e}d{\'e}rale de Lausanne (EPFL), CH-1015 Lausanne, Switzerland.  Qian Wang: Ecole Polytechnique F{\'e}d{\'e}rale de Lausanne (EPFL), CH-1015 Lausanne, Switzerland. M. Hossein Gorji: Laboratory of Multiscale Studies in Building Physics, Empa,  Swiss Federal Laboratories for Materials Science and Technology, D\"{u}bendorf, Switzerland}
\date{November 1, 2021}

\usepackage{algorithm2e}
\usepackage{algorithmic}
\usepackage{amsmath,bm}
\usepackage{subcaption}
\definecolor{ms}{rgb}{0, 0, 0}
\definecolor{ms_rev}{rgb}{0, 0, 0}
\definecolor{ms_rev1}{rgb}{0, 0, 0}

\DeclareMathOperator\erf{erf}

\usepackage{tikz,everypage}
\AtBeginDocument{%
  \AddEverypageHook{%
    \begin{tikzpicture}[remember picture,overlay]
      \path (current page.north west) --  (current page.south west) \foreach \i in {1,...,\fakelinenos} { node [pos={(\i-.5)/\fakelinenos}, xshift=\fakelinenoshift, line number style] {\i} }  ;
    \end{tikzpicture}%
  }%
}
\tikzset{%
  line numbers/.store in=\fakelinenos,
  line numbers=50,
  line number shift/.store in=\fakelinenoshift,
  line number shift=5mm,
  line number style/.style={text=gray},
}

\begin{document}

\begin{abstract}
An important class of multi-scale flow scenarios deals with an interplay between kinetic and continuum phenomena. While hybrid solvers provide a natural way to cope with these settings, two issues {\color{ms}restrict} their performance. Foremost, the inverse problem implied by estimating distributions has to be addressed, to provide boundary conditions for the kinetic solver. The next issue comes from defining a robust yet accurate switching criterion between the two solvers. This study introduces a data-driven kinetic-continuum coupling, where the Maximum-Entropy-Distribution (MED) is employed to parametrize distributions arising from continuum field variables. Two regression methodologies of Gaussian-Processes (GPs) and Artificial-Neural-Networks (ANNs) are utilized to predict MEDs efficiently. Hence the MED estimates are employed to carry out the coupling, besides providing a switching criterion. To achieve the latter, a continuum breakdown parameter is defined by means of the Fisher information distance computed from the MED estimates. We test the performance of our devised MED estimators by recovering bi-modal densities. Next, MED estimates are integrated into a hybrid kinetic-continuum solution algorithm. Here Direct Simulation Monte-Carlo (DSMC) and Smoothed-Particle Hydrodynamics (SPH) are chosen as kinetic and continuum solvers, respectively. The problem of monatomic gas inside Sod's shock tube is investigated, where DSMC-SPH coupling is realized by applying the devised MED estimates. Very good agreements with respect to benchmark solutions along with a promising speed-up are observed in our reported test cases.
\end{abstract}

\maketitle
\section{Introduction}
\label{sec:intro}
\noindent It is often the case that in real{\color{ms}-}world flow phenomena, the underlying thermo-fluid processes cover a wide range of scales. In particular, gas dynamic problems may encounter a huge variation of the Knudsen number, resulting in a failure of common numerical approaches. While the Navier-Stokes-Fourier (NSF) system {\color{ms}of equations} provides an accurate description of the flow at the hydrodynamic limit, they fail when significant departures from the equilibrium are encountered. On the other hand, the kinetic framework becomes more relevant once the breakdown of the continuum is observed. In particular, the Boltzmann equation offers a high fidelity governing equation for {\color{ms} the} evolution of the molecular distribution function. Even though the statistical description provided by the Boltzmann equation holds accurate in the whole range of the Knudsen number, numerical deficiencies have to be dealt with near the continuum. Either using Direct Simulation Monte-Carlo (DSMC) \cite{Bird1963,bird1994molecular,gallis2009convergence}, discrete velocity methods \cite{broadwell_1964,mieussens2000discrete,morris2011monte} or spectral discretizations \cite{Pareschi2000,wu2013deterministic,gamba2009spectral}, the efficiency of resulting simulations significantly drops as the collision integral becomes dense. While simplified kinetic models such as Fokker-Planck type approximations \cite{jenny2010solution,gorji2011fokker,mathiaud2016fokker,singh2015fokker} have been proposed to improve the efficiency in the near continuum and early transitional regimes, still the resulting models are not computationally competitive with respect to the more mature continuum solvers. These all motivate coupling of {\color{ms} the} continuum and kinetic solvers in order to tackle a larger set of flow problems, where both continuum and kinetic scales are present. Examples include but are not limited to unconventional gas reservoirs \cite{Sander2017}, re-entry problem \cite{tseng2006simulations}, Stratospheric aerosols \cite{kremser2016stratospheric}, high -pressure shock tubes \cite{Petersen2001}, Sonoluminescence \cite{brenner2002single}, and liquid-vapor interface \cite{DAHMS20131667}. 
\\ \ \\
An extensive set of hybrid algorithms to address the kinetic-continuum coupling has been devised in past decades. Hybrid numerical schemes have been proposed which operate on particles \cite{tiwari2009particle,burt2009hybrid,ye2012multi,gorji2015fokker}, Partial Differential Equations (PDEs) \cite{schwartzentruber2006hybrid,wang2003hybrid,aktas2002combined} or distributions \cite{xu2010unified}. While each approach aims at providing universal accuracy and efficiency, two fundamental issues are common themes in {\color{ms}the} majority of the algorithms. {\color{ms}The f}irst issue comes from the fact that kinetic and continuum equations deal with different types of variables. The variable of interest is the distribution in the former, whereas the latter relies on its moments. Clearly, extracting moments from a given distribution is a straightforward task. Yet the inverse{\color{ms},} known as the \textit{moment problem}, is not well-posed in general. Common recipes to address the moment problem in hybrid approaches, include Chapman-Enskog (CE) \cite{Chapman1953,garcia1998generation} or Grad-Hermite distributions \cite{struchtrup2003regularization,grad1949kinetic}. The second issue arises due to the demand for a relevant switching criterion which preserves {\color{ms} the} efficiency and accuracy of the underlying schemes. Typically, some norm of non-equilibrium moments or their gradients are employed to provide an optimal criterion for the breakdown of the continuum \cite{wang2003predicting,bird1970breakdown}. \\ \ \\
Adopting CE \cite{tiwari2009particle} or Grad-Hermite distributions \cite{di2016lattice}, comes with a risk of running to distributions with negative parts, which leads to the breakdown of the notion of entropy. On the other hand, the Maximum-Entropy-Distribution (MED) can provide a well-defined probability density while honoring a given set of moments. Furthermore, it can be shown that MED leads to the least-biased estimation of the distribution subject to the given moments. It is important to note that having a well-defined distribution in hand, one can also compute relevant distances of the distribution from the equilibrium to obtain a switching criterion. \\ \ \\
By virtue of the method of Lagrange multipliers, it is {\color{ms}straightforward} to show that MED takes a closed form of the exponential function \cite{kapur1989maximum}. Therefore, the moment problem reduces to finding the Lagrange multipliers for a given set of moments. Although the resulting optimization problem is well-posed in bounded domains subject to realizable moments \cite{dreyer1987maximisation,levermore1996moment}, the iterative gradient descent-type algorithms have to be employed. This comes with a significant cost since the Hessian matrix of the objective function can become ill-conditioned. Hence, alternatives such as adaptive basis methods have been pursued \cite{abramov2007improved,abramov2009multidimensional}.\\ \ \\     
A different perspective is realized by resetting the MED problem into the regression framework. More precisely, the idea becomes to compute the Lagrange multipliers as a map from input moments using regression schemes.  Recently,  a data-driven approach was introduced by the authors \cite{sadr2019GPRMED}, where Gaussian-Processes (GPs) have been adopted to provide a regression map for estimating MEDs. Either using GP regressions or Artificial Neural Networks (ANN), first, the model is trained over a data-set that fills relevant subsets of the moment space. Then, efficient estimations of MDEs are obtained for a given input of moments.
\\ \ \\
In this study, after training GP and ANN for a MED data-set, a hybrid kinetic-continuum solution algorithm based on efficient MED estimators is devised. 
Using efficient estimates of MEDs and by choosing a threshold, then a criterion for the breakdown of the continuum can be found, arising as a distance between the estimated MED and the equilibrium. Therefore based on the computed distance, either continuum or kinetic solver is adopted for a given computational cell and time step. 
\\ \ \\
To narrow down our context, we address a hybrid particle-particle solution algorithm, where the NSF system is solved using Smoothed-Particle Hydrodynamics (SPH) \cite{liu2003smoothed}, and DSMC is employed for the Boltzmann equation. However{\color{ms},} note that the methodologies presented here, are quite general and can be applied to a wider class of multi-scale solvers. Once computational cells are switched from SPH to DSMC, the DSMC particles need to be sampled from the estimated MED. Although more advanced sampling techniques such as slice sampling \cite{neal2003slice} and adaptive rejection sampling \cite{gilks1992adaptive} could be utilized, for simplicity we rely on the Metropolis-Hastings algorithm \cite{chib1995understanding} for generating DSMC particles. On the other hand, when DSMC cells convert to the SPH ones, the SPH particles are generated based on moments obtained from the DSMC particles.
\\ \ \\
The content of this manuscript is distributed among the following sections. First, the NSF system of equations{\color{ms},} along with the Boltzmann equation{\color{ms}, is} reviewed in \S~\ref{sec:NSF_Boltz}, to the extent needed in our study. Then in \S~\ref{sec:rev_methods}, SPH and DSMC solution algorithms required for solving the adopted governing equations are considered. Next in \S~\ref{sec:coup_scales}, the basic framework of the hybrid DSMC-SPH algorithm is discussed. Afterward{\color{ms},} in \S~\ref{sec:MED}, the Maximum-Entropy method and the regression approaches based on GP and ANN are devised. A hybrid solution algorithm is then proposed in \S~\ref{sec:hyb_cont_kin_part_part_alg}, where a data-driven coupling based on our MED estimator equipped with the switching criterion {\color{ms_rev}is} adopted. In order to evaluate the regression machinery, the trained MED estimators based on GP and ANN are first examined by estimating a bi-modal distribution. Then the devised hybrid solution algorithm is deployed to solve the well-known Sod's shock tube problem in \S~\ref{sec:results}. In the end, a conclusion along with a projection for next studies {\color{ms}is} presented in \S~\ref{sec:conc_disc_outlook}. 

\section{Governing equations}
\label{sec:NSF_Boltz}
\noindent Before proceed to the solution algorithms and data-driven coupling, let us recall the governing equations relevant for our continuum and kinetic settings.
\subsection{Navier-Stokes-Fourier system {\color{ms}of equations}}
\label{sec:NSF}
\noindent The NSF equations provide a system of conservation laws for evolution of the density $\rho\ {\color{ms_rev}\in \mathbb{R}}$, the momentum $\rho \bm U$ {\color{ms_rev}with $\bm U\in \mathbb{R}^3$}, and the total energy $E:=\frac{1}{2} \rho \bm U\cdot \bm U+\rho c_v T$. Here $c_v:={3 k_b}/{(2 m)}$ indicates the heat capacity of a monatomic gas at constant volume, $k_b$ is the Boltzmann constant, $T{\color{ms_rev}\in \mathbb{R}}$ is the temperature, and $m$ is the molecular mass. A general form of the conservation laws can be cast to 
\begin{flalign}
\label{eq:mass}
    &\frac{\partial \rho}{\partial t} 
+ \frac{\partial }{\partial x_i}(\rho U_i)
= 0,\\
    &\dfrac{\partial (\rho U_i) }{\partial t} 
+
 \dfrac{\partial  }{\partial x_j} (\rho  U_i U_j+ p_{ij}) = 0 \ \ \ \  \\
\text{and} \ \ \ \ \ \ &\dfrac{\partial E }{\partial t}+\dfrac{\partial}{\partial x_i}\left(EU_i+q_i  + p_{ik} U_k \right)=0~,
\label{eq:energy}
\end{flalign}
where $\bm p\ {\color{ms_rev}\in \mathbb{R}^{3\times 3}}$ stands for the pressure tensor and $\bm q\ {\color{ms_rev}\in \mathbb{R}^3}$ denotes the heat flux vector. Note that Einstein's summation notation is used throughout this study{\color{ms_rev}, i.e. $a_i b_i = \sum_i a_i b_i$}, and $\delta_{ij}$ denotes the Kronecker delta. By assuming the ideal gas law besides the Stokes closure, we get  $p_{ij} = \pi_{ij} + p \delta_{ij}$, where
\begin{flalign}
p&=n k_b T\\
\textrm{and} \ \ \ \ \ \pi_{ij} &= -2\mu S_{ij}
\end{flalign}
with $n$ the number density, $\mu$ the viscosity and
\begin{flalign}
S_{ij}=\frac{1}{2}\left( \frac{\partial U_i}{\partial x_j} + \frac{\partial U_j}{\partial x_i} \right) - \frac{1}{3}  \frac{\partial U_k}{\partial x_k}\delta_{ij}.
\end{flalign}
Moreover, the heat flux follows the Fourier law
\begin{flalign}
q_i &= - \kappa \frac{\partial T}{\partial x_i}~,
\end{flalign}
where $\kappa$ denotes the heat conductivity. For further details on the NSF equations along with the corresponding derivations, see e.g. \cite{anderson1992governing}.
\\ \ \\
\noindent As the fluid experiences strong departures from the equilibrium, it is well established that the NSF system fails to provide an accurate physical description. This is due to the fact that closure assumptions applied to the pressure tensor and heat fluxes may not hold precise once the Knudsen number becomes large. This failure is pronounced in many non-equilibrium settings including cold-to-hot heat fluxes in the lid-driven cavity \cite{rana2013robust}, slip velocity \cite{gu2009high}, shock profile \cite{timokhin2017different}, and inverted temperature gradients \cite{frezzotti2003evidence}.  \\ \ \\
To cope with these shortcomings, a higher level of closure is necessary. The kinetic theory provides a mesoscale closure by considering the dynamics of the molecular velocity distribution, which is discussed in the following.
\subsection{The Boltzmann equation}
\label{sec:Boltz}
\noindent  The statistical account of an ideal monatomic gas can be fully described by the probability density $f(\bm V|\bm x,t)$, which gives the probability $f(\bm V|\bm x,t)d^3\bm V$ for finding a particle with a velocity in the vicinity of $\bm V$, at a given position $\bm x$ and instant in time $t$. For convenience, let us consider the mass density function $\mathcal{F}(\bm V,\bm x,t):=\rho(\bm x,t) f(\bm V|\bm x,t)$ (MDF). Assuming that the gas is \textit{dilute} and the \textit{molecular chaos} holds \cite{Chapman1953}, the evolution of  MDF follows the Boltzmann equation
\begin{flalign}
&\frac{\partial \mathcal{F}}{\partial t}  + \frac{\partial (\mathcal{F} V_j)}{\partial x_j} = 
S^{\text{Boltz.}}(\mathcal{F}) 
\label{eq:Boltz_eq}
\end{flalign}
with the Boltzmann collision operator
\begin{flalign}
&S^{\text{Boltz.}}(\mathcal{F})
= \int_{\mathbb{R}^3} \int_{0}^{2\pi} \int_{0}^{+\infty}  
 \Big [  \mathcal{F}({\bm V}^*, {\bm x})\mathcal{F}({\bm V}^*_1,{\bm x})
 -\mathcal{F}({\bm V}, {\bm x})\mathcal{F}({\bm V}_1, {\bm x}) \Big ] 
g {\color{ms_rev}\hat{b}\  d\hat{b}\  d\hat{\epsilon}\ d^3  {\bm V}_1}.
\end{flalign}
The collision operator acts as a source term in the evolution of $\mathcal{F}(\bm V,\bm x,t)$, when a particle with velocity $\bm V$ is produced as the outcome of colliding particles with velocities $({\bm V}^*,{\bm V}_1^*)$, and as a sink term once one of the colliding pair  has the pre-collision velocity of $\bm V$. Note that the magnitude of the relative velocity is indicated by $g = |\bm V - \bm V_1|$. Furthermore, the collision plane is determined with the impact parameter $\hat{b}$ and the scattering angle $\hat{\epsilon}$.
\\ \ \\
Once the Boltzmann equation is solved, the macroscopic properties such as $\rho$, $\bm U$, $\bm p$, $T$ and $\bm q$ can be easily obtained by taking the moments of MDF. Since{\color{ms},} in contrast to the NSF system, here no assumption on the distance of $\mathcal{F}$ from the equilibrium is adopted, the Boltzmann equation can {\color{ms}also be employed} for flow scenarios far from the equilibrium.
\section{Particle solution algorithms}
\label{sec:rev_methods}
\noindent There exists a {\color{ms}broad} set of numerical schemes to tackle flow phenomena governed by the NSF system as well as the Boltzmann equation. However{\color{ms},} here we focus on particle approaches, mainly because the coupling of two particle systems is technically more convenient. Therefore first, a particle method based on SPH as the solution algorithm for the continuum scale is reviewed. Then, DSMC is explained as our method of choice for the Boltzmann equation. At the end of each subsection, the algorithms used in this study are summarized explicitly.
\subsection{Smoothed-Particle Hydrodynamics}
\label{sec:rev_SPH}
\noindent As a particle algorithm for solving the NSF system, the main idea behind SPH relies on notions of the weight $m_{\textrm{SPH}}$ and the kernel $W(r,h)$ carried by each particle \cite{monaghan1992smoothed,liu2003smoothed}. Here $r$ is the distance and $h$ the smoothing parameter. Let the superscript $(.)^{(\alpha)}$ denote the quantity evaluated at the location of the particle $\alpha$, and $\tilde{(.)}^{(\alpha)}$ the corresponding approximation introduced by the scheme. Furthermore $W^{(\alpha\beta)}_h=W(r^{(\alpha\beta)},h)$ is the value of the kernel evaluated at $r^{(\alpha \beta)} := | {\color{ms_rev} \bm x}^{(\alpha)} -   {\color{ms_rev} \bm x}^{(\beta)}|$ which is the distance between particles with index $\alpha$ and $\beta$. The density thus can be estimated by
\begin{flalign}
\tilde{\rho}^{(\alpha)} = \sum_{\beta} {m^{(\beta)}_{\textrm{SPH}}} W^{(\alpha \beta)}~.
\label{eq:SPH_rho}
\end{flalign}
Next, a quantity of interest $A$ at ${\bm x}^{(\alpha)}$ is estimated by 
\begin{flalign}
\tilde{A}^{(\alpha)} = \sum_{\beta} \frac{{m^{(\beta)}_{\textrm{SPH}}}}{\tilde{\rho}^{(\beta)}} {\tilde{A}^{(\beta)}} W^{(\alpha \beta)}.
\label{eq:SPH_A}
\end{flalign}
\noindent The accuracy and efficiency of SPH method depend on the choice of the kernel function and many such interpolation kernels are introduced in the literature (see e.g. \cite{monaghan2005smoothed,sigalotti2006shock}). For simplicity here, we adopt the Gaussian kernel which takes the following form in the one-dimensional space
\begin{flalign}
W(r^{(\alpha \beta)}, h) = \frac{1}{{\sqrt{2\pi} h}}{\exp\left(-\dfrac{r^{(\alpha \beta)}}{2h^2}\right)}~.
\end{flalign}
\noindent Typically a fixed small $h$ and a large cut-off  $r_{\text{cut}}$ are used. Here they are set based on the grid size $\Delta x$ via
\begin{flalign}
r_\text{cut} &= 2 \Delta x  \\
\text{and} \ \ \ \ h &= \frac{1}{30}{r_{\text{cut}}}~.
\end{flalign}
\noindent One of the main advantages of the SPH framework is the smooth estimation of the derivatives which are required in the governing equations. It is easy to see that the derivative of $A$ at the position of a particle with index $\alpha$ becomes
\begin{flalign}
\nabla  \tilde{A}^{(\alpha)} 
= \sum_{\beta} \frac{{m^{(\beta)}_{\textrm{SPH}}}}{\tilde{\rho}^{(\beta)}} {\tilde{A}^{(\beta)}}
 \nabla W^{(\alpha \beta)}~.
\end{flalign}
\noindent Therefore, by considering the Lagrangian framework for the NSF system of equations, the forces and heat transfer experienced by each particle can be computed by a time-integration scheme. 
We adopt the following discretization of momentum and energy equations
\begin{flalign}
\frac{D}{Dt} \tilde{U}_i^{(\alpha)} &=  -\sum_{\beta} \frac{{m^{(\beta)}_{\textrm{SPH}}}}{{\tilde{\rho}^{(\alpha)}}{\tilde{\rho}^{(\beta)}}}\left({\tilde{p}_{ij}^{(\alpha)}}+ {\tilde{p}_{ij}^{(\beta)}}\right) \frac{\partial W^{(\alpha\beta)}}{\partial x_j^{(\beta)}},
\label{eq:SPH_velocity}
\\
\textrm{and} \ \ \ \ \ \frac{D }{Dt}{\tilde{e}}^{(\alpha)} &= - \frac{1}{2}\sum_{\beta} \frac{{m^{(\beta)}_{\textrm{SPH}}}}{{\tilde{\rho}^{(\alpha)}}{\tilde{\rho}^{(\beta)}}} \left\{\left({{\tilde{p}}^{(\alpha)}}+ {{{\tilde{p}}^{(\beta)}}}\right) {\tilde{U}}_{i}^{( \alpha\beta)}+(\kappa^{(\alpha)}+\kappa^{(\beta)}){\tilde{T}}^{( \alpha\beta)}\right\} \frac{\partial W^{(\alpha\beta)}}{\partial x_i^{(\beta)}}
\nonumber\\ &+ \frac{2\mu^{(\alpha)}}{{\tilde{\rho}}^{(\alpha)}} {{\tilde{S}}_{ij}}^{(\alpha)} {{\tilde{S}}_{ij}}^{(\alpha)},
\label{eq:SPH_energy}
\end{flalign}
\noindent where $e=c_vT$ is the internal energy,  $U_{i}^{( \alpha\beta)}= U_{i}^{( \alpha)}-U_{i}^{(\beta)}$ and $T^{(\alpha\beta)}=T^{(\alpha)}-T^{(\beta)}$. For details of derivations see \cite{liu2003smoothed,monaghan2005smoothed}.
\noindent Adopting a first-order explicit time integration, the particles evolve in the solution domain accordingly. After initializing particles at every cell, the particles evolve according to Algorithm~\ref{alg:SPH}.
\\ \ \\
\begin{algorithm}[H]
\SetAlgoLined
 \While{$t<T_{\color{ms}\mathrm{final}}$}{
  -Compute density of each particle from \eqref{eq:SPH_rho}\;
  -Estimate  $\bm S$ and $\bm p$ for every particle\;
  -Evolve velocity and internal energy of each particle \eqref{eq:SPH_velocity}-\eqref{eq:SPH_energy}\;
  -Stream particles with their velocities\;
  -Increment $t$\;
 }
 \caption{SPH solution algorithm for the NSF system}
 \label{alg:SPH}
\end{algorithm}
\subsection{Direct Simulation Monte-Carlo}
\label{sec:rev_DSMC}
\noindent Due to the high-dimensionality of the solution space associate with MDF, particle Monte-Carlo schemes become attractive for numerical solutions of the Boltzmann equation. Consider an ensemble of particles with velocities $\bm M^{{\color{ms}(}i{\color{ms})}}$, positions $\bm X^{{\color{ms}(}i{\color{ms})}}$ and weights $w^{{\color{ms}(}i{\color{ms})}}$. The MDF $\mathcal{F}(\bm V,\bm x, t)$ is related to particles states through 
\begin{flalign}
\label{eq:MDF-identity}
\mathcal{F}(\bm V,\bm x,t)&=\sum_i \delta(\bm X^{{\color{ms}(}i{\color{ms})}}-\bm x)\delta(\bm M^{{\color{ms}(}i{\color{ms})}}-\bm V)w^{{\color{ms}(}i{\color{ms})}},
\end{flalign}
where $\delta(.)$ is the Dirac delta
 \cite{rjasanow2005stochastic,pfeiffer2015two,gorji2014efficient}.
{\color{ms}For simplicity, we consider the same weight for all DSMC particles.}
Using the particle description, the idea behind DSMC is that instead of updating $\mathcal{F}$ according to the {\color{ms_rev}Boltzmann Eq.~\eqref{eq:Boltz_eq}}, the underlying jump process is simulated. Hence, the evolution of positions $\bm X^{{\color{ms}(}i{\color{ms})}}$ and velocities $\bm M^{{\color{ms}(}i{\color{ms})}}$ are derived consistent with the Boltzmann equation in a two-step manner: streaming and collision. While the streaming phase is simply the free flight of particles, the collision follows the No-Time-Counter (NTC) method of Bird \cite{bird1994molecular}. The collision probability is found via 
\begin{flalign}
\mathcal{P}_{\text{coll.}} = \frac{\sigma_T c_r}{(\sigma_T c_r)_\text{max}},
\label{eq:col_prob}
\end{flalign}
where $\sigma_T$ indicates the collision cross-section, $c_r{\color{ms_rev}=|\bm c_r|}$ is the magnitude of the relative velocity for a colliding pair, and the subscript ${(.)}_{\max}$ indicates the maximum value. In case of the Hard-Sphere molecular potential, the collision cross-section is $\sigma_T = \pi \sigma^2$ where $\sigma $ is the diameter of the molecules.
Collisions occur for the colliding pair $(i,j)$ in a manner that guarantees the conservation of mass, momentum, and energy. Let $\bm c_c=(\bm M^{{\color{ms}(}i{\color{ms})}}+\bm M^{{\color{ms}(}j{\color{ms})}})/2$ be the center of mass velocity and $\bm c_r=(\bm M^{{\color{ms}(}i{\color{ms})}}-\bm M^{{\color{ms}(}j{\color{ms})}})$ the relative velocity. According to the Hard-Sphere scattering law, the post-collision relative velocity is isotropic. Hence the angles
\begin{flalign}
\theta &= \mathrm{arccos}( 2 \alpha_1-1 ) \label{eq:dsmc_coll_1} \ \  \\
\textrm{and} \ \ \ \phi &= 2 \pi \alpha_2
\label{eq:dsmc_coll_2}
\end{flalign}
with uniformly distributed random numbers $\alpha_{1,2}\in [0,1]$, result in the orientation of the post-collision relative velocity
\begin{flalign}
{\bm c}^*_r &=  c_r \bigg(\cos(\theta), \ \sin(\theta)\cos(\phi), \ \sin(\theta)\sin(\phi)\bigg)^T. \label{eq:dsmc_coll_3}
\end{flalign}
The post-collision particles velocities are fully determined by relative and center of mass velocities, following
\begin{flalign}
{\bm M}^{{\color{ms}(}i{\color{ms})}} &=   \bm c_{c} + \frac{{\bm c}^*_r}{2}  \label{eq:dsmc_coll_4}\\
\textrm{and} \ \ \ \ {\bm M}^{{\color{ms}(}j{\color{ms})}} &=   \bm c_{c} - \frac{{\bm c}^*_r}{2}~.
\label{eq:dsmc_coll_5}
\end{flalign}
\sloppy A short description of DSMC algorithm for Hard-Spheres, after initializing the particles is summarized in Algorithm~\ref{alg:DSMC}. {\color{ms_rev} Note that in each computational cell of volume $V_\mathrm{cell}$, the collision probability given by Eq.~\eqref{eq:col_prob} provides the cap on the number of colliding-pair candidates  $N_\mathrm{Cand}=\frac{1}{2} N_{\mathrm{p}/\mathrm{cell}}^2 F_N (\sigma_T c_r)_{\mathrm{max}} \Delta t/V_{\mathrm{cell}}$, where $N_{p/\mathrm{cell}}$ is the number of particles per cell with identical statistical weight $F_N$ during one time step $\Delta t$.}
\begin{algorithm}
\SetAlgoLined
 \While{$t<T_\mathrm{final}$}{
  -Sample the moments\;
  \For{$i=1,...,N_{\mathrm{cells}}$}{
    -$N_{\mathrm{Cand}} = \frac{1}{2} N_{\mathrm{p}/\mathrm{cell}}^2 F_N (\sigma_T c_r)_{\mathrm{max}} \Delta t/V_{\mathrm{cell}}$\;
    \For{$j=1,...,N_{\mathrm{Cand}}$}{
    -Pick two samples from the cell\; 
    -Draw a random number $r$ with a uniform distribution in $[0,1]$\;
    \If{$\sigma_T c_r/(\sigma_T c_r)_{\mathrm{max}}<r$}{
    -Perform the collision \eqref{eq:dsmc_coll_1}-\eqref{eq:dsmc_coll_5}\;
    }
    }
  }
  -Stream position of particles $\bm X$ with their velocities $\bm M$\;
  -Increment $t$\;
 }
 \caption{DSMC algorithm for Hard-Spheres. $F_N{\color{ms}:=w/m}$ indicates the statistical weight, $V_{\mathrm{cell}}$ is the volume of the computational cell, $N_{\mathrm{p/cell}}$ is the number of particles per cell, and $N_{\textrm{Cand}}$ is number of candidates to be considered for collisions at each time step.}
 \label{alg:DSMC}
\end{algorithm}
\section{Upscaling and refinement}
\label{sec:coup_scales}
\noindent We focus on a domain decomposition coupling approach, where each of {\color{ms}the} DSMC/SPH solvers will be employed in the designated sub-domains. Therefore two types of communications between continuum and kinetic solvers have to be addressed. First{\color{ms},} we discuss moment recovery which is upscaling of the kinetic information to the continuum scale. Next, we address the moment problem arising from {\color{ms}the} refinement of the continuum information into the kinetic scale. The former gives us the information passage from kinetic to continuum sub-domains, whereas the latter accounts for the opposite{\color{ms}.}
\subsection{Moment estimations}
\label{sec:kin_to_cont}
\noindent The evolution of moments can be easily obtained from the Boltzmann equation by taking the velocity moments of Eq.~\eqref{eq:Boltz_eq}. By  comparing the flux terms with respect to the conservation laws i.e. Eqs.~\eqref{eq:mass}-\eqref{eq:energy}, the corresponding macroscopic quantities are readily obtained 
\begin{flalign}
U_i&=\frac{1}{\rho} \int_{\mathbb{R}^3} V_i \mathcal{F} d^3 \bm V \\
p_{ij} &=  \int_{\mathbb{R}^3} \xi_{ i} \xi_{j } \mathcal{F} d^3 \bm V \\
\textrm{and} \ \ \ \ q_i &=\frac{1}{2} \int_{\mathbb{R}^3} \xi_{ i} \xi_{j } \xi_{j } \mathcal{F} d^3 \bm V, 
\label{eq:heat_flux}
\end{flalign}
where $ \bm \xi :=  \bm V -  \bm U$ denotes the fluctuating velocity. {\color{ms_rev} Please note the Einstein summation notation in $j$ index for the heat flux in  Eq.~\eqref{eq:heat_flux}.} By applying the identity \eqref{eq:MDF-identity}, the above-mentioned moments can be computed from DSMC particles. Yet since a finite number of particles and finite size of computational cells are applied in DSMC, the corresponding equalities turn to estimates; assigned here by the superscript $\tilde{(.)}$. The field variables at position $\bm x$ and time $t$ are found according to
\begin{flalign}
\tilde{\rho}&=\frac{1}{\delta \Gamma}\sum_{\bm k \in \mathcal{S}_\Gamma} w^k, \\
\tilde{U}_i&= \frac{1}{\tilde{\rho}}\sum_{\bm k \in \mathcal{S}_\Gamma} M_i^k w^k, \\
\tilde{p}_{ij}&= \sum_{k \in \mathcal{S}_\Gamma} {M_i^\prime}^k {M_j^\prime}^kw^k  \\
\textrm{and} \ \ \ \ \tilde{q}_i&= \frac{1}{2}\sum_{k \in \mathcal{S}_\Gamma} {M_i^\prime}^k {M_j^\prime}^k{M_j^\prime}^k w^k, 
\end{flalign}
where $\mathcal{S}_\Gamma$ is the set of particles residing in the computational cell $\Gamma$ around {\color{ms}position} $\bm x$ at time $t$, and $\delta \Gamma$ denotes the volume of that cell. Furthermore $\bm M^\prime:=\bm M-{\color{ms}\tilde{\bm U}}$ is the fluctuating particle velocity. The above estimates commit two types of errors. First, the statistical noise arising from using a limited number of particles and second, the spatial homogenization due to the spatial averaging over the computational cell. Note that the former error can be significantly reduced once the time averaging can be utilized relevant for the stationary flows.
\subsection{Moment problem}
\label{sec:cont_to_kin}
\noindent The more challenging part of the coupling is when the information provided by the continuum solver should be transferred to the kinetic sub-domain. In other words, a distribution should be assigned to the moments given by the continuum model. Note that in general this moment problem is ill-posed and needs a regularization for tractability. There exist many approaches addressing the moment problem, including the quadrature method \cite{fox2009higher}, Grad-Hermite distributions \cite{torrilhon2016modeling,Torrilhon2013} and Maximum-Entropy methods \cite{levermore1996moment,dreyer1987maximisation}. Assuming realizable moments, the quadrature method gives a well-defined probability density, yet suffers from non-uniqueness of the distribution. On the other hand, the Grad-Hermite ansatz gives a unique solution which can be computed cheaply, while it can not guarantee positivity of the probability density \cite{Torrilhon2013}. Finally MED gives rise to a unique and positive probability density, providing a bounded domain and realizable set of moments \cite{hauck2008convex}. 
From the information-theory stand-point, MED provides the least-biased distribution subject to the moment constraints.
In the following section, we first review some basics of the Maximum-Entropy approach. Furthermore, we explain and elaborate the recently devised data-driven method as an efficient way to compute MED probability densities \cite{sadr2019GPRMED}.
\section{Maximum-Entropy Distribution}
\label{sec:MED}
\noindent For simplicity but without loss of generality, in the following we focus on the one-dimensional velocity space. Let us consider finding a probability density that fulfils the moments
\begin{flalign}
\label{eq:constraint}
\int_\Omega f {\bm \phi} dV &={\bm p},
\end{flalign}
where $ {\bm p}=(\rho, \rho U, p, ...)^T$ is the vector of given moments and $ {\bm \phi} = (1, V, \xi^2, ...)^T$ the corresponding vector of velocity polynomials. 
The Maximum-Entropy approach provides a regularization of the moment problem by minimizing the entropy functional
\begin{flalign}
S[f] = \int_\Omega f \ln(f) dV~,
\end{flalign}
subject to the moment contraints \eqref{eq:constraint}.
\noindent As shown in \cite{hauck2008convex}, the resulting optimization problem is well-posed on a bounded domain $\Omega$ and for realizable moments $\bm p$. Yet note that in case of an unbounded domain, such distribution still uniquely exists once further restrictions on  higher order moments are imposed \cite{noonan1976inverse,einbu1977existence}. Following the method of Lagrange multipliers, the objective functional becomes
\begin{align}
    C^{ \lambda}_N [f] := 
    \int_\Omega f \ln(f) d V
    - \lambda_k\left( \int_\Omega f \phi_k d  V- p_k \right)
    \label{eq:obj_max_entropy}
\end{align}
where $\bm \lambda$ is vector of Lagrange multipliers owing to the moment constraints. By taking the variational derivative of $C_N^\lambda$, it is easy to see that the  extremum of the functional becomes
\begin{align}
    f_N^{ \lambda}  = Z_\lambda^{-1}\ {\exp{\left(-\lambda_k \phi_k\right)}} ~,
    \label{eq:MED}
\end{align}
where the denominator $Z_\lambda := \int_\Omega \exp{\left(-\lambda_k \phi_k\right)} dV$ normalizes the MED $f_N^{\lambda}$ \cite{kapur1989maximum}. Note that at this stage the moment problem is reduced to finding the vector $\bm \lambda$.
Next, considering the dual problem, we arrive at the following unconstrained minimization problem
\begin{flalign}
\bm \lambda(\bm p)=\argmin_{\lambda^*\in \mathbb{R}^N}\left\{Z_{\lambda^*}-\lambda^*_j p_j\right\},
\label{eq:dual}
\end{flalign}
which can be directly employed in order to find the Lagrange multipliers $\bm \lambda$.
\subsection{Direct method}
\label{sec:direct}
\noindent
The common approach for solving the minimization problem given in Eq.~\eqref{eq:dual}, follows Newton's method.  The resulting iterative solution algorithm updates the estimate ${\bm \lambda}^n$ by ${\bm \lambda}^{n+1}$, according to
\begin{align}
  &H_{ij}(\bm \lambda^n) \Delta { \lambda^n_j} =  g_i(\bm \lambda^n) \ \  \\
 \text{and} \ \ \ \  &\lambda^{n+1}_i=\lambda^{n}_i+\beta^n\Delta \lambda^n_i~.
\end{align}
Here $\beta^n$ is a damping factor at $n$th iteration, and {\color{ms_rev}$\bm H(\bm \lambda)$ and $  \bm g(\bm \lambda)$} are the Hessian and the gradient of the objective function, respectively. Since the Hessian matrix is a function of the previous Lagrange multipliers estimate, there is no guarantee that {\color{ms_rev}$\bm H$}  is well-conditioned, hence high computational cost can be expected \cite{schaerer201735}. Alternatively, an adaptive basis provides a mechanism to avoid the ill-conditioning issue. While the change of the basis, e.g., using Hermit polynomials to enforce a close to diagonal Hessian \cite{schaerer201735}, or devising a Hessian for the intermediate probability density at each iteration \cite{abramov2007improved,abramov2009multidimensional}, can tackle the  problem with regards to ill-conditioned Hessian, the overhead computational cost still constraints the use of direct methods.
\subsection{Gaussian-Process Regression}
\noindent Following \cite{sadr2019GPRMED}, instead of directly solving for the Lagrange multipliers as explained in \S~\ref{sec:direct}, numerical advantages can be obtained by considering a regression approach. The idea is to approximate the unique map $\Psi_i: \bm p \rightarrow \lambda_i, \ i=1,...,N$ by GP via
\begin{flalign}
\label{eq:GP-general}
\tilde{\Psi}_i \sim \mathcal{GP}(0,\mathcal{K}_i)
\end{flalign}
given a positive semi-definite kernel function $ \mathcal{\bm K}(\bm p,\bm p^\prime)$. Note that here and henceforth, $A\sim B$ reads $A$ is a sample drawn from $B$. For details of Gaussian-Process regressions see e.g. \cite{Rasmussen2006,owhadi2019operator}.
The kernel introduces a set of hyper-parameters $\Theta_i$ to be fitted to the data, which can be optimized by maximizing the log-likelihood 
\begin{align}
    \ln & \left[   \tilde{f}\left (\tilde{\Psi}_i(\bm p)\,|\,   (\bm p,\lambda_i)\in \mathcal{D}_i \right)\right]~
\end{align}
over the input/output data-set $D_i=\{ (\bm p,\lambda_i)_k \}_{k=1}^{N_{\text{data}}}$, where $\tilde{f}\left (\tilde{\Psi}_i(\bm p)\,|\,   (\bm p,\lambda_i)\in \mathcal{D}_i \right)$ denotes the probability density of $\tilde{\Psi}_i$ conditional on the training data. \\ \ \\
Once the hyper-parameters of the chosen kernel are found, the Lagrange multipliers can be predicted based on the values of input moments $\bm p^*$ according to
\begin{align}
\left(\tilde{\Psi}_i ({{\bm p}^*}) \bigg \vert  \tilde{\Psi}_i(\bm p)=\Psi_i(\bm p)\right)
\sim
\mathcal{N}(\bar{m}_i,\bar{\Sigma}_i),
\end{align}
where
\begin{align}
\bar{m}_i &= \mathcal{K}_{\Theta_{(i)}}( {{\bm p}^*}, {{\bm p}^{\prime}})   \mathcal{K}_{\Theta_{(i)}}( \bm p, {{\bm p}^{\prime}})^{-1} \Psi_{(i)}(\bm p)
\ \ \ \ \text{and}
\label{eq:pred_mean}
\\
\bar{\Sigma}_i&=\mathcal{K}_{\Theta_{i}}({\bm p}^*, {\bm p}^{*\prime}) -   \mathcal{K}_{\Theta_{(i)}}( {\bm p}^*, \bm p) \mathcal{K}_{\Theta_{(i)}}( \bm p, {\bm p}^{\prime})^{-1}  \mathcal{K}_{\Theta_{(i)}}( {\bm p}^*, \bm p)~.
\label{eq:pred_var}
\end{align}
Note that $\mathcal{N}( A, B)$ denotes the normal distribution with mean $ A$ and variance $ B$. Also note that the Einstein summation over repeated indices is abandoned if the brackets $(.)$ are used over the index.
Similar to \cite{sadr2019GPRMED}, a kernel based on the Gaussian radial basis function
\begin{flalign}
\mathcal{K}_{\Theta_i}(\bm p,{\bm p}^\prime) = \sigma_{(i)} \exp\left(- r_{(i)}^2/2 \right) + \epsilon \delta(\bm p-{\bm p}^\prime) 
\end{flalign}{}
\ \\
\noindent is used in this work, where $\epsilon \delta(\bm p-{\bm p}^\prime)$ with $\epsilon=10^{-6}$ guarantees that the outcome covariance matrix  is diagonally dominant and invertible. Moreover $r_i$ indicates an Euclidean scaled distance from the other input point. The Broyden-Fletcher-Goldfarb-Shanno algorithm (BFGS) is employed to optimize the hyperparameters as the training step of our GP model \cite{nocedal2006numerical}. Since here only a prototype study is considered, a simple GP regression using GPflow \cite{GPflow2017} is employed for estimating the Lagrange multipliers. Yet since the cost of Cholesky's factorization needed for the covariance matrix is cubic with respect to the number of training data points, more elaborate GP schemes are needed for larger data-sets. In this study, $N_{\text{data}}=1000$ data points were used for training GP.
Motivated by limitations of GP and in order to get a better insight into pros and cons of data-driven approaches, we also consider regression based on ANN as discussed in the following.
\subsection{Artificial-Neural-Network}

\begin{figure}
\centering
	\includegraphics[scale=0.9]{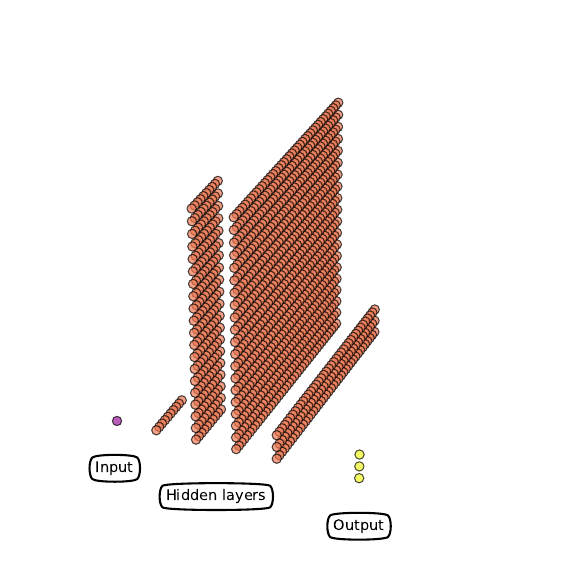}
	\caption{A schematic representation of the devised ANN for the case  $\mathrm{dim}(\lambda)=3$}
	\label{fig:neural_network}
\end{figure}

\noindent As an alternative data-driven approach and universal function approximator, ANN is employed here to address the regression problem of Lagrange multipliers. The objective is to make a comparison between ANN and GP regressions in terms of both accuracy and computational efficiency in the context of our MED problem. Therefore here we are concerned with learning the map $\Psi_i: \bm p \rightarrow \lambda_i, \ i=1,...,N$ using an ANN approximator. For further details on ANNs and their design principles see e.g. \cite{haykin1994neural,goodfellow2016deep}. \\ \ \\
We start with a small network, including only three hidden layers, and increase numbers of neurons in hidden layers $(H_1, H_2, H_3)$ until a reasonable accuracy in predicting the testing data points $(\bm p,\bm \lambda)$ is achieved. A schematic of ANN is presented in Fig.~\ref{fig:neural_network}. The forward propagation of the network with hyperbolic tangent activation function reads
\begin{flalign}
h^{{\color{ms}(}1{\color{ms})}}_i &=  \tanh(W^{{\color{ms}(}1{\color{ms})}}_{ij}p_j+b^{{\color{ms}(}1{\color{ms})}}_i),\\
h^{{\color{ms}(}2{\color{ms})}}_i &= \tanh(W^{{\color{ms}(}2{\color{ms})}}_{ij} h^{{\color{ms}(}1{\color{ms})}}_j+b^{{\color{ms}(}2{\color{ms})}}_i),\\
h^{{\color{ms}(}3{\color{ms})}}_i &= \tanh( W^{{\color{ms}(}3{\color{ms})}}_{ij} h^{{\color{ms}(}2{\color{ms})}}_j+b^{{\color{ms}(}3{\color{ms})}}_i) \\
 \text{and} \ \ \ \ {\hat{\lambda}}_i &= W^{{\color{ms}(}o{\color{ms})}}_{ij} h^{{\color{ms}(}3{\color{ms})}}_j + b^{{\color{ms}(}o{\color{ms})}}_i,
\end{flalign}
\noindent where $\bm W^{{\color{ms}(}1{\color{ms})}} \in \mathbb{R}^{H_1 \times dim(p)}$, $\ \bm b^{{\color{ms}(}1{\color{ms})}} \in \mathbb{R}^{H_1}$, $\bm W^{{\color{ms}(}2{\color{ms})}} \in \mathbb{R}^{H_2\times H_1}$, $\ \bm b^{{\color{ms}(}2{\color{ms})}} \in \mathbb{R}^{H_2}$, $\bm W^{{\color{ms}(}3{\color{ms})}} \in \mathbb{R}^{H_3\times H_2}$, $\ \bm b^{{\color{ms}(}3{\color{ms})}} \in \mathbb{R}^{H_3}$, $\bm W^{{\color{ms}(}o{\color{ms})}} \in \mathbb{R}^{dim(\lambda) \times H_3}$ and $\ \bm b^{{\color{ms}(}o{\color{ms})}} \in \mathbb{R}^{dim(\lambda)}$. Note that $\hat{\bm \lambda}\in \mathbb{R}^{\dim(\lambda)}$ indicates the ANN prediction of the Lagrange multipliers.
\noindent Coefficients of the network are found by minimizing the loss function $ \mathcal{\hat{C}}$ defined as a modified mean squared error (MSE) on the $N_{\text{tr}}$ training data points, i.e.,
 \begin{flalign}
 \mathcal{\hat{C}} {:=} \frac{1}{N_{\text{tr}}} \sum_{i=1}^{N_{\text{tr}}} || { \hat{\bm \lambda}}^{(i)} - \bm \lambda^{(i)} ||_{{\color{ms}2,}\mathbb{R}^{dim(\lambda)}}^2 
 + \sum_{i=1}^{3} \mathcal{R}(W{\color{ms}^{(i)}}) + \mathcal{R}(W{\color{ms}^{(o)}}),
 \end{flalign}{}
 \\ 
\noindent  where {$||\ .\ ||_{{\color{ms}2,}\mathbb{R}^N}$ indicates the $L^2$-norm of $N$-dimensional vector, and} $\mathcal{R}$ regulates the weights of the network to avoid over-fitting during the training step \cite{goodfellow2016deep}. In this paper, the {$L^2$} regularizer is used, corresponding to  
\begin{flalign}
\mathcal{R}(W) = \gamma ||W||^{2}_{\textrm{F}{\color{ms},\mathbb{R}^{dim(W)}}},
\end{flalign}
\noindent where subscript F indicates Frobenius norm  and the hyper-parameter $\gamma$ is used to control the regularization effect. The training of the network is implemented in Keras \cite{chollet2015keras}, with TensorFlow \cite{tensorflow2015-whitepaper} as the backend. The optimal weights and biases of the network are obtained using the Adam stochastic optimizer \cite{kingma2014adam}, which uses mini-batches of size $N_\mathrm{batch} <N_{\mathrm{tr}}$ of the training data to take a single optimization step by minimizing the loss function. More precisely, the full training data set with $N_{\mathrm{tr}}$ data-points is shuffled, and $N_{\mathrm{tr}}/N_\mathrm{batch}$ mini-batches are extracted to take $N_{\mathrm{tr}}/N_\mathrm{batch}$ optimization steps. Once the entire training data set is exhausted, the training is said to complete one training epoch. The training is performed for a sufficient number of epochs to obtain a converged network. The convergence speed of the training is controlled by the learning rate of $\eta$. A feature scaling technique in which all the components of the input are scaled to the same range is applied to the data-sets to accelerate the training process \cite{ioffe2015batch}. An input scalar $x$ is scaled by the mean normalization 
 \begin{flalign}
 x^* = \frac{x-\mu}{\sigma}~,
 \end{flalign}
\noindent where $\mu$ is the mean and $\sigma$ is the standard deviation of $x$. At the beginning of the training, the weights and biases of the network are randomly initialized using normal distributions \cite{glorot2010understanding}. Therefore, the training needs to be performed several times, following a multiple restarts approach \cite{hsu1995artificial}, to prevent the training results from depending on the initialization of the weights. In this paper, ten restarts are performed for the training of the network, and the trained model with the best validation accuracy is selected as the final model. The validation accuracy metric is the mean squared error (MSE).
\noindent The optimal network in this paper is obtained with the hyper-parameters: $(H_1,H_2,H_3)=(10,20,40)$, $\gamma=10^{-8}$, $N_\mathrm{batch}=1000$, and $\eta=0.005$.

\subsection{Generating training data-set}
\noindent Either applying GP regression or ANN approximator, we need to first produce a relevant data-set containing the areas of interest for the pair $(\bm p, \bm \lambda)$. In the context of this study we only consider regressions on MEDs with one-dimensional normalized velocity domain. The corresponding $\mathbb{R}$-domain hence is truncated to $\Omega=[-10,10]\sigma$, where $\sigma^2$ is the input variance. Since in practice mean and variance of any distribution can be modified by scaling and shifting, the data-set is generated on the distributions with mean zero and unity variance. Although data generation is more efficient by randomly generating Lagrange multipliers and then taking the integrals \cite{sadr2019GPRMED}, here we would like to have more control over the domain of the moments. Therefore in this study, the training data-set is generated by sampling from the space of moments and then directly solving for the Lagrange multipliers of MED, as explained in \S~\ref{sec:direct}. Let $\mathcal{L}_i$ be the subspace from which the moment $p_i$ is sampled. We only focus on MEDs with maximum number of eight moments. These moments are uniformly sampled according to $\mathcal{L}_3=[-2,2]$, $\mathcal{L}_4=[-5,5]$, $\mathcal{L}_6=\mathcal{L}_5=[-10,10]$ and $\mathcal{L}_8=\mathcal{L}_7=[-20,20]$.

\subsection{Training cost of GP and ANN}
\begin{figure}
  \centering
    \includegraphics[scale=1.0]{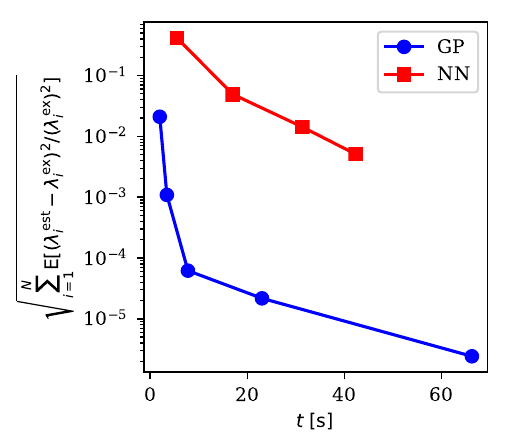}
  \caption{\color{ms_rev} Relative error in predicting MED with $N=3$ Lagrange multipliers against training time $t$ of GP and NN  using  $\{ 100, 500, 800, 1000, 1500 \}$ and $\{ 16000,32000,64000,90000 \}$ training data points, respectively. The computational time indicates the averaged time over 10 samples per test case.}
  \label{fig:cost_GP_ANN}
\end{figure}

\noindent The training cost of GP/ANN estimators with $N=3$ Lagrange multipliers is presented, with respect to the error in recovering the testing set. As shown in Fig.~\ref{fig:cost_GP_ANN}, although GP provides a better accuracy with fewer training points, its  cost increases almost cubically with the number of training points due to the computational complexity associated with inverting the covariance matrix. However, the cost of ANN scales linearly with the number of training points; yet large pool of data is required for a reasonable accuracy. Therefore, if high dimensional MED with large space of Lagrange multipliers is considered, ANN estimator would have an edge, while the use of GP for similar accuracy can be restrained by the computational resources.

\section{Hybrid DSMC-SPH solution algorithm}
\label{sec:hyb_cont_kin_part_part_alg}
\noindent In this section, the explained data-driven MED machinery is integrated into a DSMC-SPH hybrid algorithm. First, a switching criterion based on MED estimates is proposed to dynamically assign DSMC/SPH simulator to each computational cell. Having trained an efficient MED estimator, one can approximate the distance from the equilibrium. Furthermore, MED estimates provide a reasonable closure needed at the DSMC-SPH interface; giving the information fed into DSMC from SPH cells. \\ \ \\
Before proceed to further details, note that since we are concerned with DSMC/SPH solvers, only the MED with three moments will be relevant. Furthermore, we employ a normalized vector of central moments
\begin{flalign}
\hat{\bm p} = 
\left( 0, 1,
\frac{2q}{\rho (k_bT/m )^{3/2}}
\right)^T
\end{flalign}
as the input to MED estimators. Notice that since we have only presented the case with a one-dimensional velocity for our MED estimators, only one-dimensional non-equilibrium flow will be considered. Let $V_1$ be the coordinate along the non-equilibrium dimension, therefore the main assumption here is that the MDF can be decomposed as
\begin{flalign}
\label{eq:1d}
\mathcal{F}(\bm V,\bm x,t)&=\rho(\bm x,t)f(V_1|\bm x,t)f^{\mathrm{eq.}}(V_2|\bm x,t)f^{\mathrm{eq.}}(V_3|\bm x,t),
\end{flalign}
where $f^{\mathrm{eq.}}$ is a Maxwellian with the local mean velocity $\bm U(\bm x,t)$ and the local temperature $T(\bm x,t)$. 
\subsection{Switching}
{\color{ms_rev} 
\noindent The breakdown of continuum may be measured either based on gradients of macroscopic properties in the physical space (e.g. in the form of gradient-length Knudsen number \cite{schwartzentruber2006hybrid}) or via a distance from the equilibrium in the probability space. While the former can be computed in a rather straight-forward way, the latter has (at least) the conceptual superiority. For example if we consider relaxation of a non-equilibrium gas state in a physically homogeneous setting, the gradient based estimates may not be relevant (since gradients are zero). However, a breakdown based on the distance from the equilibrium provides a consistent switching criterion as far as an accurate estimate of the distribution is available. Similar to the moment problem, one can rely on MED as the least biased solution of the moment problem, to estimate the distance from the equilibrium. 
}
\ \\
\noindent A relevant metric to obtain the distance between two densities $f_1(V)$ and $f_2(V)$ is given by the Fisher information distance {\color{ms_rev}(relative Fisher information)}
\begin{flalign}
\mathcal{I}(f_1|f_2):=\int_{\mathbb{R}^3} f_1 \left[\nabla \ln({f_1}/{f_2})\right]^2 d^3 \bm V.
\label{eq:fisher}
\end{flalign}
 {\color{ms_rev}The Fisher information arises in many contexts related to the probability measures. For a detailed survey on physical importance of the Fisher information see \cite{frieden2000physics}. In Fokker-Planck type kinetic models, it can be shown that the entropy evolution is governed by the relative Fisher information \cite{gorji2020entropic}. Besides such interesting theoretical properties, the Fisher information is motivated here since the normalizer of MED, i.e., $Z_\lambda$ in Eq.~\ref{eq:MED}, disappears due to the gradient of log-pdf on $f_2$; once MED is considered as the second pdf. Hence, unlike other measures of differences between distributions, we would not need to store/compute the normalizer.} \\ \ \\
In this study, we adopt the Fisher information distance as a measure to see how far the distribution stands from the equilibrium. 
\noindent First $\hat{\bm p}$ is estimated in every cell leading to an approximate $f_3^\lambda$ based on GP and ANN regressions. Then, the switching criterion is found from $\mathcal{I}(f^{\mathrm{eq.}}|f_3^\lambda)$
which has an analytical form. By setting a tolerance $\epsilon$, the SPH cells are switched to DSMC if $\mathcal{I}(f^{\mathrm{eq.}}|f_3^\lambda) \ge \epsilon$, and vice-versa if $\mathcal{I}(f^{\mathrm{eq.}}|f_3^\lambda) < \epsilon$.\\ \ \\
To switch a continuum cell to the kinetic one, all SPH particles are removed, and new DSMC particles are sampled from $f_3^\lambda$ and positioned in the cell randomly with a uniform distribution. Conversely, DSMC particles are removed and SPH particles with values based on the moments of DSMC particles are placed equidistantly in the cell.

\subsection{Treatment of interface}
\label{sec:treat_interface}

\noindent Let us decompose the domain into two parts; the union of continuum cells with SPH particles in them, and the union of kinetic cells occupied by DSMC particles. At the interface between DSMC and SPH cells, a proper boundary condition for each method needs to be derived. In this section, two candidate numerical approaches for interface treatment, with  different efficiency/simplicity trade-off will be presented.

\subsubsection{Flux method}
\label{sec:flux_cond}
\noindent One way to treat the interface boundary problem is based on the incoming flux. In particular, we consider the flux of incoming particles, as if the neighbouring cells outside of the considered sub-domain were operating with a solver similar to the one employed inside of the sub-domain. This leads to an accurate boundary condition for each scale, although the implementation in higher dimensions can become cumbersome.
\\ \ \\
\noindent For each kinetic cell lying at the boundary, the incoming flux of the neighbour distribution can be determined from the MED estimation of the distribution at the interface. Hence, first $f_3^\lambda$ and accordingly the MDF $\mathcal{F}_3^\lambda$ are estimated from moments of the neighbouring continuum cell. Then new particles are generated based on the MDF
\begin{flalign}
\mathcal{F}_{\color{ms}\mathrm{in}}&=H(V_ln^i_l)\ V_jn^i_j\mathcal{F}_3^\lambda
\end{flalign}
of incoming particles \cite{bird1994molecular}. Note that here $H(\ .\ )$ denotes the Heaviside function and $\bm n^i$ the normal of the interface inward to the kinetic cell. The particles are sampled using Metropolis-Hasting sampling method \cite{chib1995understanding}. The new DSMC particles are streamed in the kinetic domain starting from the interface position with a uniformly distributed fraction of time step, i.e., $\delta t=r \Delta t$ where $r$ is a random number in $(0,1)$.  
\\ \ \\
For the continuum cells at the border with the kinetic ones, the flux is approximated using the moments of the neighbouring kinetic cell. Consider the mean velocity, the temperature and the {\color{ms}density} {\color{ms}estimated} at the neighbouring DSMC cell to be ${\bm U}^{{\color{ms}(}b{\color{ms})}}$, $T^{{\color{ms}(}b{\color{ms})}}$ and ${\color{ms}\rho^{(b)}}$. The incoming flux to the SPH domain is non-zero only if $U^{{\color{ms}(}b{\color{ms})}}_j\tilde{n}^i_j>0$ where $\tilde{\bm n}^i$ is the normal of the interface inward to the continuum cell. Once this condition is met, we consider the volume ${\bm U}^{{\color{ms}(}b{\color{ms})}}\Delta t$ at the neighbouring DSMC cell. Let us call the resulting domain containing that volume, $\Omega_{\color{ms}\mathrm{in}}$. The idea is to populate $\Omega_{\color{ms}\mathrm{in}}$ with SPH particles as described in the following. Based on the weights of SPH particles $w$ and ${\color{ms}\rho^{(b)}}$, new SPH particles with equidistant spacing, the velocity $\bm U^{{\color{ms}(}b{\color{ms})}}$ and the temperature $T^{{\color{ms}(}b{\color{ms})}}$ are introduced inside the $\Omega_{\color{ms}\mathrm{in}}$ sub-domain.
\\ \ \\
Although describing inflow fluxes leads to accurate and efficient treatment of sub-domain boundaries, it requires special care when it is implemented in higher dimensions. Hence an equivalent but simpler approach is suggested next, where the idea of using ghost cells for each sub-domain is explored.
\subsubsection{Ghost cells method}
\label{sec:ghost_cell}

\begin{figure}
\centering
\begin{subfigure}{0.4\columnwidth}
	\includegraphics[scale=0.9]{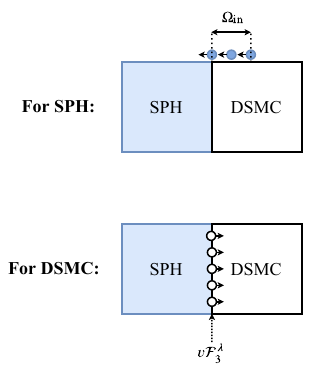}
	\caption{Flux method}
\end{subfigure}
\begin{subfigure}{0.4\columnwidth}
\includegraphics[scale=0.9]{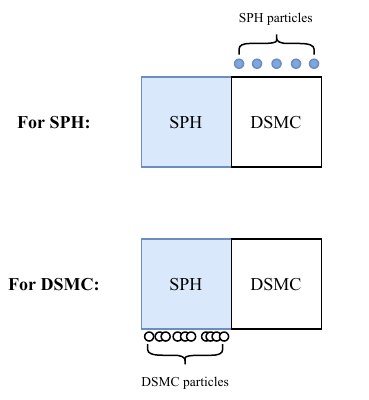}
	\caption{Ghost cell method}
\end{subfigure}
  \caption{Schematics of interface treatment}
  \label{fig:interface}
\end{figure}

\noindent From the implementation point of view, a simpler alternative can be obtained by extending each sub-domain with ghost cells. 
\\ \ \\
In the case of DSMC, at every time step and every interface, a ghost DSMC cell on top of the adjacent SPH cell is considered. Next, new DSMC particles are generated based on the MED estimate $\mathcal{F}_3^\lambda$ given the set of central moments coming from the SPH solution. Using the Metropolis-Hasting approach new velocities are assigned and particles are located in the ghost cell with a uniform distribution. Once the particles in the ghost cells are streamed, the remaining DSMC particles in the ghost cells are removed.
\\ \ \\
A similar approach can be deployed for treatment of SPH cells at the interface. On top of each neighbouring DSMC cell, a ghost cell is introduced and equidistant SPH particles are generated according to the moments of the local DSMC particles. Once SPH particles are streamed with the velocities obtained from the moments of the beneath DSMC cell, consistent fluxes are obtained. At the end of each time step, the SPH particles remaining in ghost cells are removed.  \\ \ \\
The above-mentioned approaches to cope with the boundary conditions for DSMC/SPH sub-domains are depicted schematically in Fig.~\ref{fig:interface}. {\color{ms_rev} Theoretically both approaches are consistent, as the two methods provide the same distribution of particles that cross the interface. From a practical aspect, no major difference was observed in our validation studies. For our simulation runs, the flux method was deployed as the method of choice for the interface treatment.}

\subsubsection{Conservation across interface}

\begin{figure}
\centering
	\includegraphics[scale=0.9]{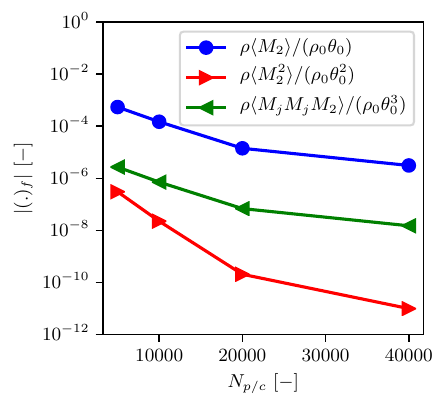}
	\caption{{\color{ms_rev} Total flux computed at the interface based on ghost cell approach \S~\ref{sec:ghost_cell} against number of particles per cell $N_{p/c}$ for fluid flow at the equilibrium. Here, we simulated the stationary fluid at equilibrium inside a box with specular walls, imposed a SPH-DSMC decomposition in physical space and turned off the switching mechanism. Mass, momentum and energy fluxes are normalized with the initial fluid density $\rho_0$ and  thermal velocity $\theta_0=\sqrt{k_b T_0/m}$ where $T_0$ indicates the initial temperature.}}
	\label{fig:cons_across_interface}
\end{figure}

\noindent Interface treatments introduced here are expected to satisfy the conservation laws subject to a statistical noise. To check this in practice, we consider the fluid inside an equilibrium box where SPH-DSMC decomposition in physical space is imposed. Since the flow is at the equilibrium we expect no fluxes of mass, momentum or energy across the interface. In  Fig.~\ref{fig:cons_across_interface}, the numerical results are reported where the deviation from equilibrium values are found to be only due to limited number of statistical particles. 

\subsection{Algorithm}
\label{sec:hyb_alg}

\noindent By assembling all the introduced pieces into a hybrid DSMC-SPH particle method, the solution algorithm is summarized in Algorithm~\ref{sec:hyb_alg}.  Note that while only one-dimensional case is considered here, all methodologies are extendable to higher dimensions in a straightforward manner.
\ \\ \ \\
The algorithm is initialized by the continuum model everywhere in the solution domain, where discretization with the SPH method is done via initializing equidistant particles in each cell with equal weight, velocity and temperature. During the simulation time evolution, the MED is estimated for every cell in the domain through GP/ANN estimator for the normalized moments obtained from the SPH particles. As soon as the estimated MED of one cell departs significantly from the Maxwellian, i.e., $\mathcal{I}(f^{\mathrm{eq.}}|f_3^\lambda)>\epsilon$ for a given tolerance $\epsilon$,
the SPH cell converts to DSMC.  %
The relative Fisher information between the two densities is  computed  here analytically in the bounded one-diemnsional domain
\begin{flalign}
\int_{-a}^a f^\mathrm{eq.}[\nabla \ln( f^\mathrm{eq.}/f^\lambda_3)]^2 dV =& \sqrt{2 \pi} \erf{\frac{a}{\sqrt{2}}} (\lambda_1^2 + 6 \lambda_1 \lambda_3 + 4 \lambda_2^2 - 4 \lambda_2 + 27 \lambda_3^2 + 1)
\\
&- 2 a e^{-{a^2}/{2}} (9 a^2 \lambda_3^2 + 6 \lambda_1 \lambda_3 + 4 \lambda_2^2 - 4 \lambda_2 + 27 \lambda_2^2 + 1).
\end{flalign}
The transformation from SPH to DSMC is achieved by removing SPH particles, generating DSMC particles sampled from $f_3^\lambda$ using Metropolis-Hasting and rescaling the velocities back to the original frame by fixing the mean and variance.
\\ \ \\
As the algorithm evolves further, both DSMC and SPH solvers require appropriate inflow/outflow boundary condition as described in \S~\ref{sec:treat_interface}. On top of inflow/outflow, the  SPH algorithm requires a description of moments outside its domain at the interface since the kernel of SPH particles can be extended to DSMC cells. Therefore, we consider virtual ghost SPH particles on top of the DSMC cells adjacent to the interface,  equidistant from one another and with equal weight, velocity and temperature. The weight, velocity and temperature are updated according to moments of the DSMC solver. If the MED estimated distribution of the DSMC cell gets close to the Maxwellian, i.e., $\mathcal{I}(f^{\mathrm{eq.}}|f_3^\lambda)<\epsilon$, the DSMC cell converts back to SPH by removing DSMC particles and introducing virtual ghost SPH particles to the SPH flow.
\\ \ \\
\begin{algorithm}[H]
\SetAlgoLined
 Initialization with SPH particles in every cell\;
 \While{$t<T_\mathrm{final}$}{
  -Evolve particles states according to SPH or DSMC rules\;
  -Apply DSMC-SPH interface treatments\;
  -Remove particles in wrong cells\;
  \For{{cell}$=1,...,N_{\mathrm{cells}}$}{
  -Estimate $f_3^\lambda$ from moments using GP or ANN\;
  \If{$\mathcal{I}(f^{\mathrm{eq.}}|f_3^\lambda)>\epsilon$ and $\mathrm{cell}_\mathrm{flag}$$==$SPH}{
   -Replace SPH particles with samples of $f_3^\lambda$
   \;
   }
   \If{$\mathcal{I}(f^{\mathrm{eq.}}|f_3^\lambda)<\epsilon$ and $\mathrm{cell}_\mathrm{flag}$$==$DSMC}{
   -Remove DSMC particles\;
   
   -Generate SPH particles with local moments\;
   }
   }
  -Increment $t$\;
 }
 \caption{Hybrid DSMC-SPH solution algorithm{\color{ms_rev1}. Unless mentioned otherwise, the breakdown parameter is set to $\epsilon=8\times10^{-6}$ in this study.}}
\end{algorithm}

\section{Results}
\label{sec:results}
\noindent The  hybrid solution algorithm based on the data-driven MED estimate devised in previous sections \S~\ref{sec:coup_scales}-\ref{sec:hyb_cont_kin_part_part_alg} is examined here against the benchmark results. First, the accuracy of GP and ANN regressions to estimate MEDs is studied in \S~\ref{sec:bi_mod}. Then, in \S~\ref{sec:shock_tube}, the Sod's shock tube as a test case with evident non-equilibrium effects is simulated using DSMC, SPH and the introduced hybrid algorithm. Before proceed, note that in all MED related computations we truncate the sample space to the domain $\Omega=[-10,10]\sigma$, where $\sigma^2$ is the variance.
\subsection{Estimating a bi-modal distribution}
\label{sec:bi_mod}
\begin{figure}
  \centering
     \includegraphics[scale=0.6]{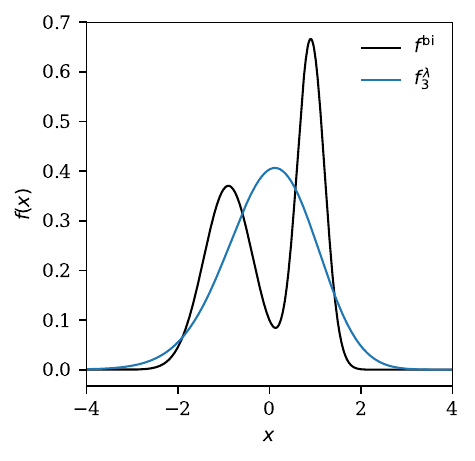}
     \includegraphics[scale=0.6]{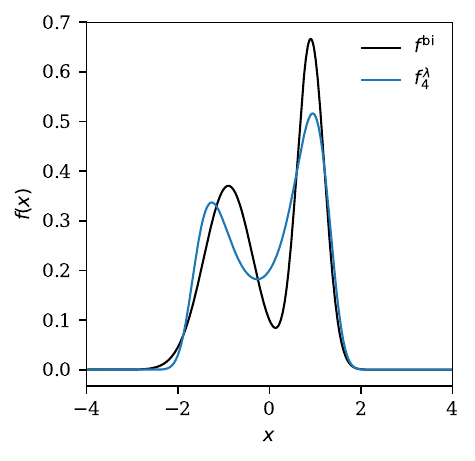}
     \includegraphics[scale=0.6]{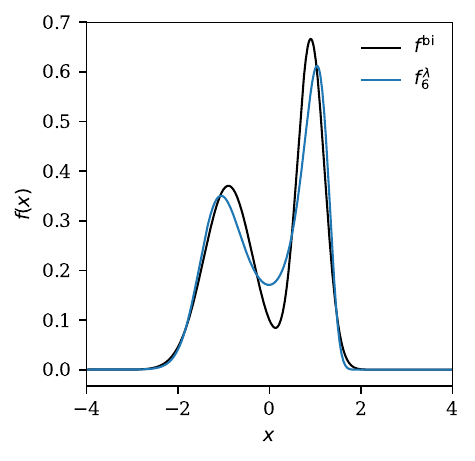}
     \includegraphics[scale=0.6]{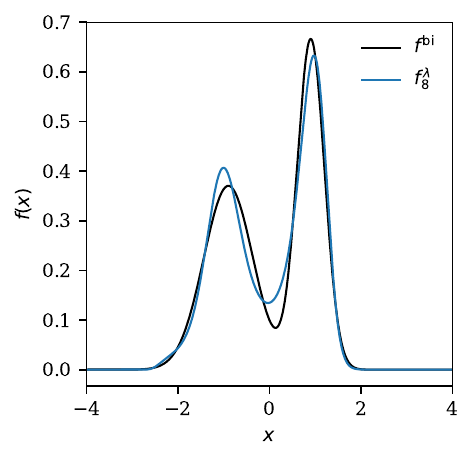}
     \includegraphics[scale=0.6]{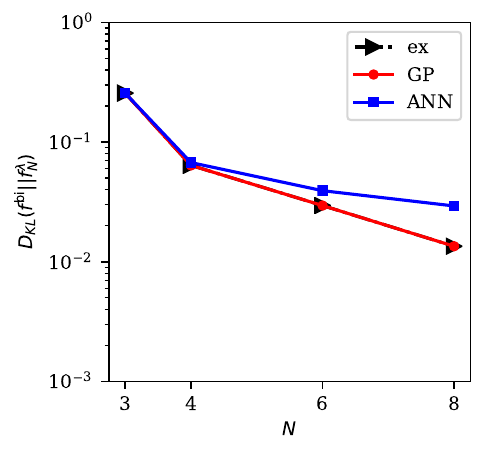}
     \includegraphics[scale=0.6]{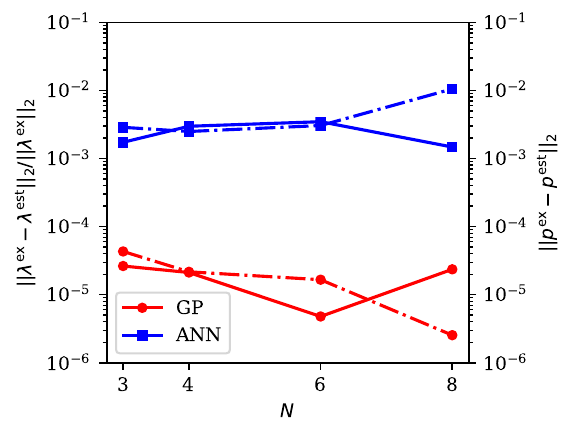}
  \caption{Estimating $f^{\mathrm{bi}}(x|\mu_1, \sigma_1, \mu_2, \sigma_2) = \frac{1}{2} \left[f^{\mathcal{N}}(x|\mu_1,\sigma_1)+f^{\mathcal{N}}(x|\mu_2,\sigma_2) \right]$ with $\mu_2 = - \mu_1$ and $\sigma_2=\sqrt{2 - (\sigma_1 ^ 2 + 2  \mu_1 ^ 2)}$ using exact, GP and ANN estimates of MED $f_N^\lambda$. Here, $\mu_1 = 0.9$ and $\sigma_1 = 0.3$. {\color{ms_rev} The MED pdfs depicted on upper panels as well as the lower left one for $N=3,4,6,8$ are the exact MED solutions of the moment problem.} For the errors, the solid and dashed lines indicate the relative error in estimating Lagrange multipliers and the error in the outcome moments of the estimated MED, respectively.}
  \label{fig:bimodal}
\end{figure}

\begin{figure}
  \centering
  \begin{minipage}[l]{0.3\textwidth}
\centering
    \includegraphics[scale=0.6]{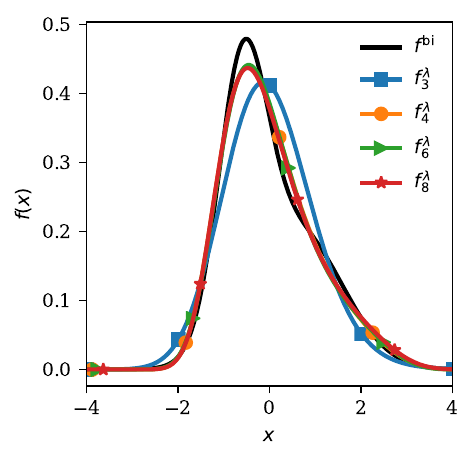}
    \includegraphics[scale=0.6]{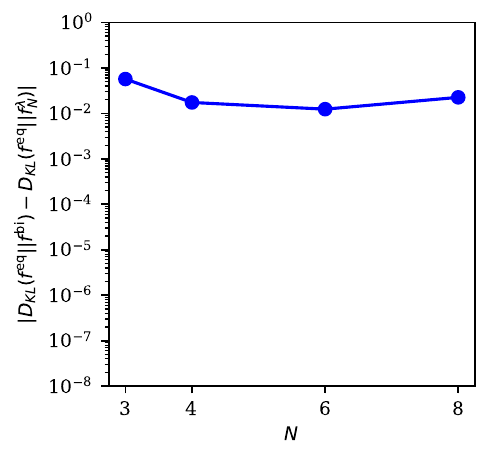}
    \subcaption{}
\end{minipage}
\begin{minipage}[c]{0.3\textwidth}
\centering
    \includegraphics[scale=0.6]{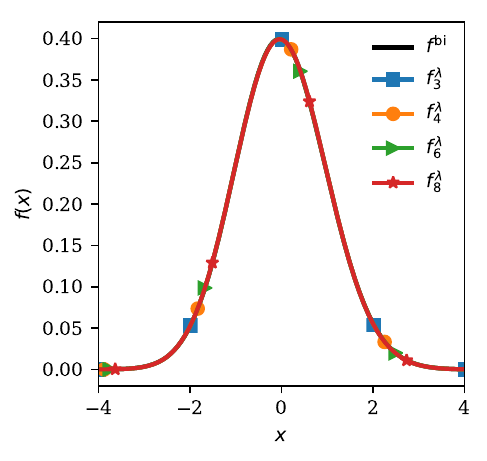}
    \includegraphics[scale=0.6]{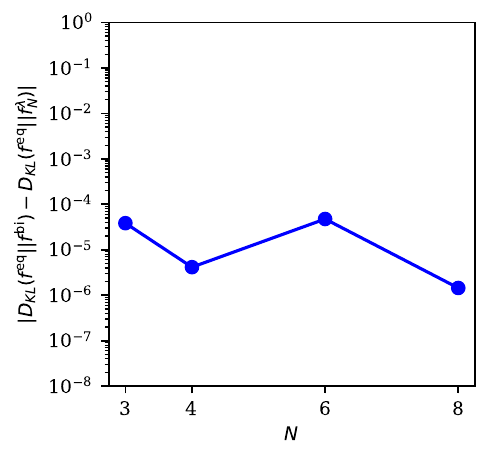}
    \subcaption{}
\end{minipage}
\begin{minipage}[r]{0.3\textwidth}
\centering
    \includegraphics[scale=0.6]{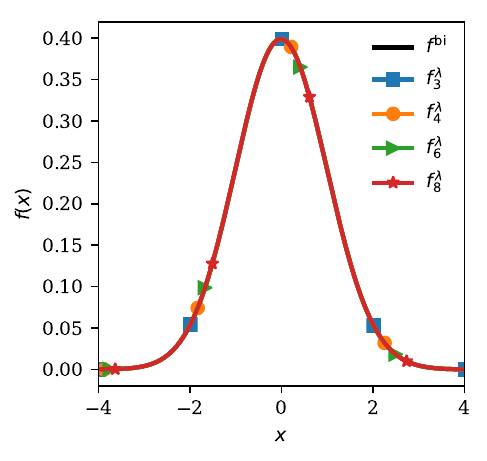}
    \includegraphics[scale=0.6]{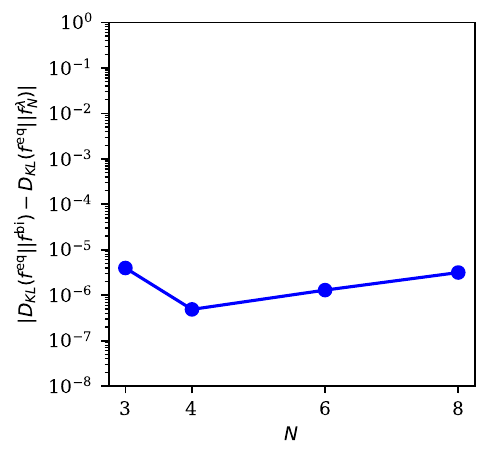}
    \subcaption{}
\end{minipage}
  \caption{\color{ms_rev} Estimating $f^{\mathrm{bi}}(x|\mu_1, \sigma_1, \mu_2, \sigma_2) = \frac{1}{2} \left[f^{\mathcal{N}}(x|\mu_1,\sigma_1)+f^{\mathcal{N}}(x|\mu_2,\sigma_2) \right]$ with $\mu_2 = - \mu_1$ and $\sigma_2=\sqrt{2 - (\sigma_1 ^ 2 + 2  \mu_1 ^ 2)}$ approaching Gaussian distribution using GP estimate of MED $f_N^\lambda$; (a) $\mu_1 = 0.6$ and $\sigma_1 = 1.0$, (b) $\mu_1 = 0.3$ and $\sigma_1 = 1.0$ and (c) $\mu_1 = 0.2$ and $\sigma_1 = 1.0$.}
  \label{fig:bimodal_eq}
\end{figure}

\begin{figure}
  \centering
  \begin{minipage}[l]{0.3\textwidth}
\centering
    \includegraphics[scale=0.6]{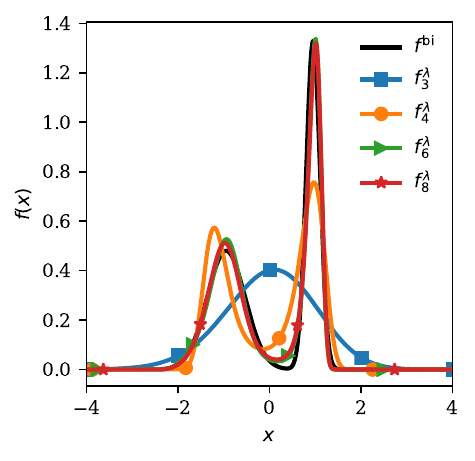}
    \includegraphics[scale=0.6]{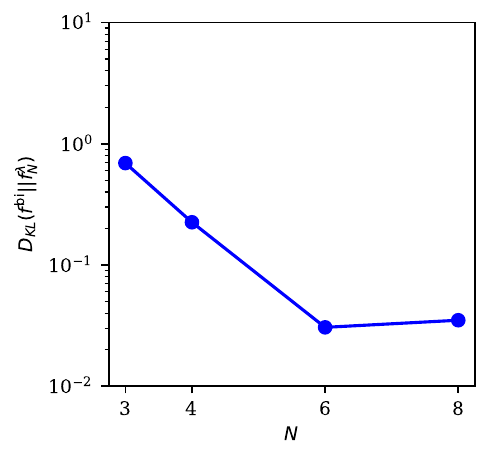}
    \subcaption{}
\end{minipage}
\begin{minipage}[c]{0.3\textwidth}
\centering
    \includegraphics[scale=0.6]{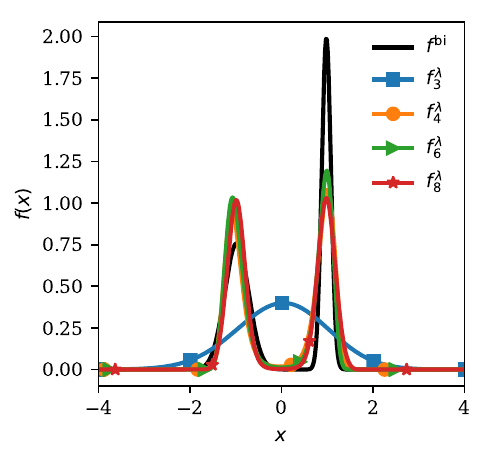}
    \includegraphics[scale=0.6]{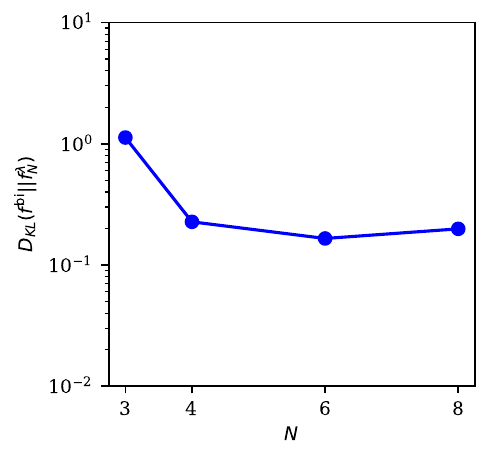}
    \subcaption{}
\end{minipage}
\begin{minipage}[r]{0.3\textwidth}
\centering
    \includegraphics[scale=0.6]{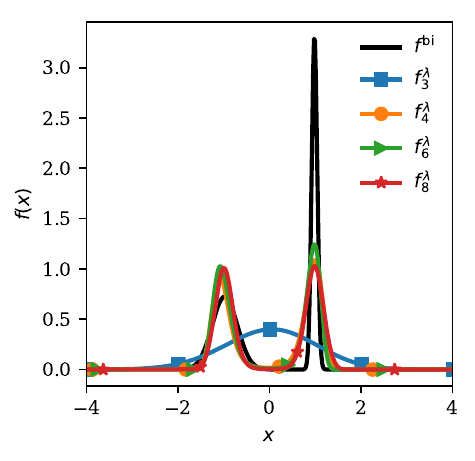}
    \includegraphics[scale=0.6]{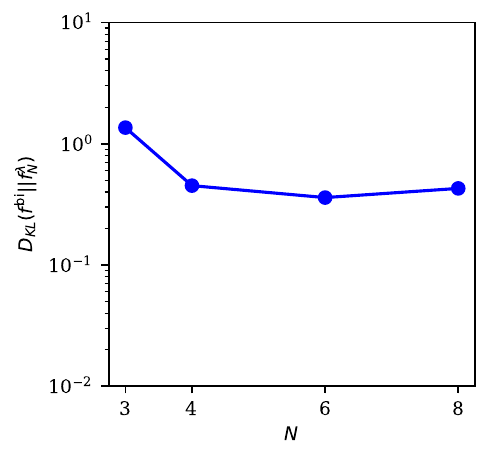}
    \subcaption{}
\end{minipage}
  \caption{\color{ms_rev1} Estimating $f^{\mathrm{bi}}(x|\mu_1, \sigma_1, \mu_2, \sigma_2) = \frac{1}{2} \left[f^{\mathcal{N}}(x|\mu_1,\sigma_1)+f^{\mathcal{N}}(x|\mu_2,\sigma_2) \right]$ with $\mu_2 = - \mu_1$ and $\sigma_2=\sqrt{2 - (\sigma_1 ^ 2 + 2  \mu_1 ^ 2)}$ approaching a dirac type distribution using GP estimate of MED $f_N^\lambda$; (a) $\mu_1 = 0.95$, $\sigma_1 = 0.15$ and $D_{KL}(f^\mathrm{eq}||f^\mathrm{bi})=1.381$, (b) $\mu_1 = 0.98$, $\sigma_1 = 0.1$ and $D_{KL}(f^\mathrm{eq}||f^\mathrm{bi})=3.429$, and (c) $\mu_1 = 0.98$, $\sigma_1 = 0.06$ and $D_{KL}(f^\mathrm{eq}||f^\mathrm{bi})=3.887$.}
  \label{fig:bimodal_far_eq}
\end{figure}

\noindent Before testing the devised hybrid multi-scale solution algorithm, first, the accuracy of the trained GP and ANN for $N=3,4,6\ \mathrm{and}\ 8$ number of moments is evaluated. The reference is the bi-modal probability density which is an archetype of non-equilibrium flows (see e.g. \cite{solovchuk2010prediction}). Consider a bi-modal density obtained by adding two Gaussian ones each of the form $f^{\mathcal{N}}(x|\mu, \sigma)$ with mean $\mu$ and standard deviation $\sigma$
\begin{flalign}
f^{\mathrm{bi}}(x|\mu_1, \sigma_1, \mu_2, \sigma_2) = 
\frac{1}{2} \left[f^{\mathcal{N}}(x|\mu_1,\sigma_1)+f^{\mathcal{N}}(x|\mu_2,\sigma_2) \right]
\end{flalign}{}

\noindent where $\mu_2 = - \mu_1$ and $\sigma_2=\sqrt{2 - (\sigma_1 ^ 2 + 2  \mu_1 ^ 2)}$ to make sure that  mean and variance of $f^{\mathrm{bi}}(x|\mu_1, \sigma_1, \mu_2, \sigma_2)$ is equal to zero and one, respectively. The moments of $f^{\mathrm{bi}}(x|\mu_1, \sigma_1, \mu_2, \sigma_2)$ can be easily obtained by taking the integral in a large enough bounded domain, numerically. Now there are two questions to be addressed here. First, given the set of moments $p$, how well MED can approximate $f^{\mathrm{bi}}$? Next, how well GP or ANN can recover the underlying MED estimate?
\\ \ \\
In order to answer the first question, the direct approach to recover MED is applied, where the converged solution with the error in moments up to the machine accuracy $\epsilon=10^{-16}$ is considered as the exact solution. Since the direct approach can be expensive if the initial guess is too far from the solution, the output from the trained GP estimate is used as the initial guess here. As shown in Fig.~\ref{fig:bimodal}, by increasing the number of moments and, consequently, the number of Lagrange multipliers, the exact MED solution converges to the bi-modal distribution. Here, the Kullback-Leibler (KL) divergence
\begin{flalign}
D_{KL} (f^{\mathrm{bi}}|| f_N^\lambda) = \int_\mathbb{R} f^{\mathrm{bi}} \textrm{ln}\left({f^{\mathrm{bi}}}/{f_N^\lambda} \right) dx
\end{flalign}
\noindent is employed as an indicator of the distance between the two densities. As it is shown in Fig.~\ref{fig:bimodal}, the KL divergence between the target and the exact MED decreases as more moments are engaged.
\\ \ \\
Having investigated the accuracy of the exact MED in recovering the bi-modal density, let us turn to the second question and evaluate the MED estimates using GP and ANN regressions.  As depicted in Fig.~\ref{fig:bimodal}, both GP and ANN provide reasonable accuracy in predicting the exact MED in terms of the KL-divergence point of view.  Yet one can observe that the GP regression seems to produce a better estimation overall considering the errors in Lagrange multipliers and the outcome moments. Unlike ANN, GP regressions provide the uncertainty of predictions in terms of the variance. In the considered test cases, the variance of prediction was in the order of $10^{-6}$, which also indicates that the prediction point is not far from the training data-set.
\\ \ \\
In terms of the computational efficiency, the cost of predictions using GP or ANN are fixed. For the former, the cost depends on the number of training data points, whereas the latter is a function of the network dimension. In this test case, the direct approach for finding MED is at least one order of magnitude more expensive than data-driven estimates. However note that in this comparison, we did not consider neither the offline costs associated with producing the data-sets, nor tuning/training GP/ANN.
{\color{ms_rev}\ \\ \ \\As our focus in this work is on MED estimates for hybridizing the continuum-kinetic solution algorithm, we further investigate the accuracy of the MED estimates around the Maxwellian distribution{\color{ms_rev1}\ as well as dirac-like distributions}. In particular, we consider three cases where the bi-modal distribution approaches the equilibrium{\color{ms_rev1}\ and three cases approaching dirac-bump in the distributions}. As it can be seen in Fig.~\ref{fig:bimodal_eq}, the error in estimating the KL divergence from the equilibrium decreases as the target distribution gets closer to the equilibrium.{\color{ms_rev1}\ Moreover, as shown in Fig.~\ref{fig:bimodal_far_eq}, the GP estimate of bi-modal distribution as the bump tends to dirac-like function deteriorates. This could be improved by including more data points near the target distribution.} This result indicates that the fitted MED estimators are more accurate near the equilibrium, which further justifies their use in the Fisher information distance switching criterion for coupling continuum model with the kinetic one.}

\subsection{Sod's shock tube}
\label{sec:shock_tube}
\begin{figure}
  \centering
\includegraphics[scale=0.9]{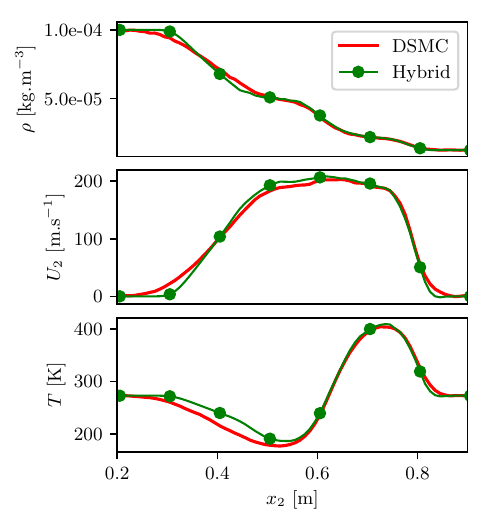}
\includegraphics[scale=0.9]{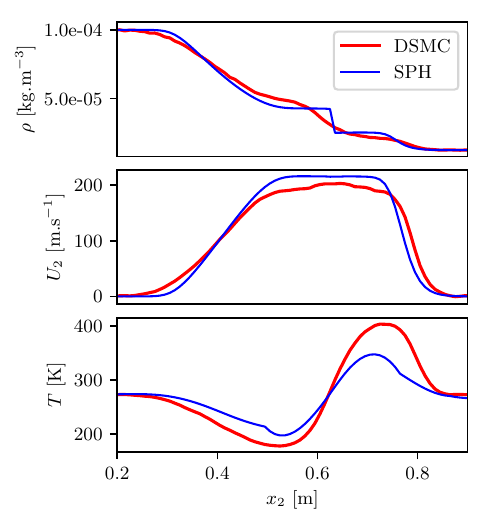}
  \caption{Sod's shock tube with initial values of $(\rho_L, U_{2,L}, T_L) = (10^{-4} ~\mathrm{kg.m^{-3}}, 0 ~\mathrm{m.s^{-1}}, 273 ~\mathrm{K})$ and $(\rho_R, U_{2,R}, T_R) = (0.125\times 10^{-4} ~\mathrm{kg.m^{-3}}, 0~\mathrm{m.s^{-1}}, 273~\mathrm{K})$ at {\color{ms_rev}$t = 8.39\times 10^{-4}\ \mathrm{s}$}. The profiles of the density $\rho$, the mean velocity $U_2$, and the temperature $T$ are obtained from SPH, DSMC, and the devised hybrid DSMC-SPH equipped with GP estimator of MED.}
  \label{fig:shock_tube_GPR_1e-4}
\end{figure}

\begin{figure}
  \centering
\includegraphics[scale=0.9]{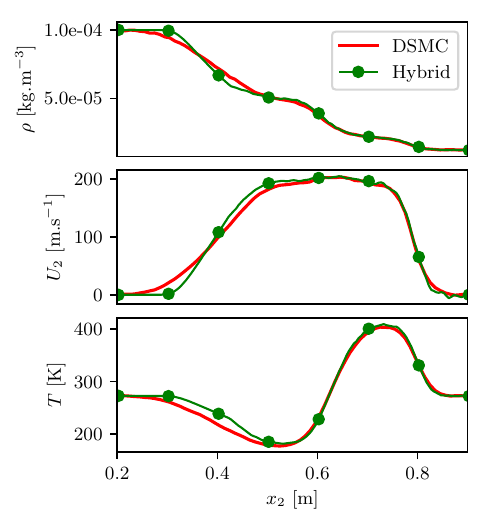}
\includegraphics[scale=0.9]{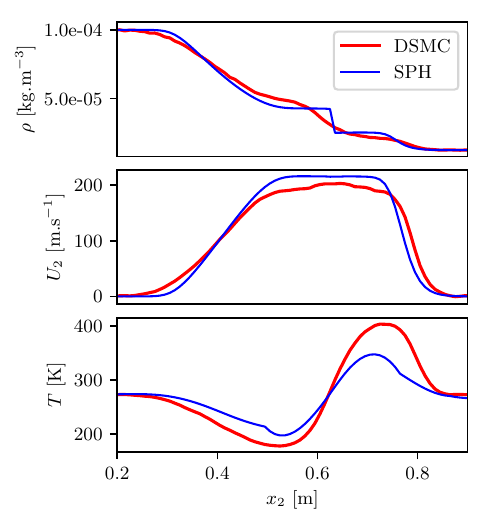}
  \caption{Sod's shock tube with initial values of $(\rho_L, U_{2,L}, T_L) = (10^{-4} ~\mathrm{kg.m^{-3}}, 0 ~\mathrm{m.s^{-1}}, 273 ~\mathrm{K})$ and $(\rho_R, U_{2,R}, T_R) = (0.125\times 10^{-4} ~\mathrm{kg.m^{-3}}, 0~\mathrm{m.s^{-1}}, 273~\mathrm{K})$ at {\color{ms_rev}$t = 8.39\times 10^{-4}\ \mathrm{s}$}. The profiles of the density $\rho$, the mean velocity $U_2$, and the temperature $T$ are obtained from SPH, DSMC, and the devised hybrid DSMC-SPH equipped with ANN estimator of MED.}
  \label{fig:shock_tube_NN_1e-4}
\end{figure}

\begin{figure}
  \centering
\includegraphics[scale=0.9]{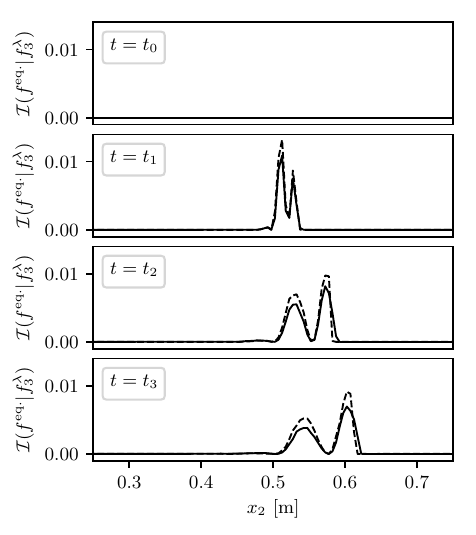}
\includegraphics[scale=0.9]{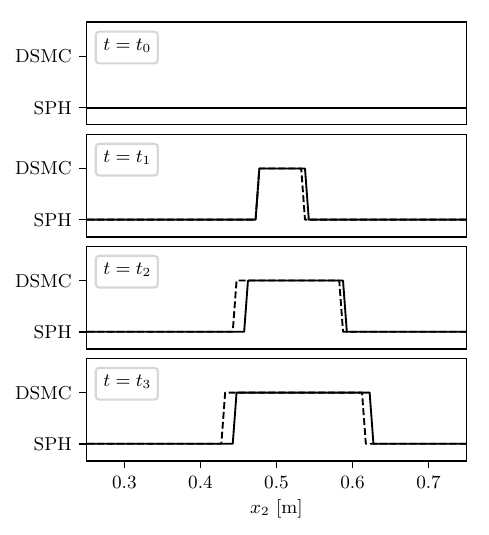}
  \caption{Evolution of the Fisher information distance $\mathcal{I}(f^{\mathrm{eq.}}|f_3^\lambda)$ at $t \in \{ 0,0.5025, 1.5075, 2.01 \}\times 10^{-4}\ \mathrm{s}$ and the assigned DSMC/SPH solver in the hybrid method for the Sod's shock tube test case with $\rho_0=10^{-4}\ \mathrm{kg.m^{-3}}$, shown at left and right, respectively. Solid and dashed lines indicate that MED is estimated using GP and ANN, respectively}
  \label{fig:shock_tube_Ifish_GPR_1e-4}
\end{figure}

\noindent In this section, the continuum-kinetic hybrid solution algorithm devised in \S~\ref{sec:hyb_alg} is applied to study of the classical Sod's shock tube \cite{sod1978survey}. Note that similar settings were used by \cite{tiwari2009particle}, to evaluate their hybrid algorithm. Consider a domain where the relevant fluid direction is in $x_2 \in (0,L)$ with $L=1\ \mathrm{m}$. The other dimensions in the physical space are assumed to be large enough, and consequently the kinetics of the gas along these dimensions can be considered to be at the equilibrium and thus condition \eqref{eq:1d} is met. For the continuum model, consider the initial values 
\begin{flalign}
\left\{\begin{matrix}
\rho_{L} = \rho_0 \ \ \\ 
U_{2,L}=u_0\ \ \\ 
T_{L}=T_0\ \ 
\end{matrix}\right.
\ \  \text{and} \ \ 
\left\{\begin{matrix}
\rho_R = \rho_0/8 \ \ \\ 
U_{2,R}=u_0\ \ \\ 
T_R=T_0\ \ 
\end{matrix}\right.
\end{flalign}
\noindent where $\rho_0 \in \{10^{-4}, 10^{-5}, 10^{-6}\}\ \mathrm{kg.m^{-3}}$, the initial temperature $T_0=273\ \mathrm{K}$, the initial velocity of $u_0=0\ \mathrm{m.s^{-1}}$, and the subscripts $L$ and $R$ indicate the left and right sides of the initial discontinuity, respectively.\\ \ \\
In the case of DSMC, while particle number follows the initial densities, the initial velocity in each direction $i\in\{1,2,3\}$ follows
\begin{flalign}
M_{i,L/R} \sim {\mathcal{N}}(U_{0},kT_{L/R}/m),
\end{flalign}
where the subscript $(.)_{L/R}$ denotes left or right side of the initial discontinuity. \\ \ \\
The Neumann boundary conditions are applied at two ends of the physical domain. In the case of DSMC, particles leaving the domain are removed while new particles entering the domain are sampled from the flux of the Maxwell distribution with moments taken from the adjacent DSMC cell. In the SPH case, as particles leave the domain, they are removed while for the inflow duplicates of the adjacent SPH cell are streamed.
\\ \ \\
After an initial convergence study, a descritization using $200$ cells leading to the mesh size of $\Delta x = 0.005\ m$ is used. The time step is picked based on $\Delta t = 0.01 \times \textrm{min}\left( t_{\text{MFT}}, \Delta x/\tilde{U} \right)$, where $\tilde{U} = \max{\left(U, \mu/\rho \right)}$, the mean-free-time is indicated by $t_{\text{MFT}}=\lambda( k_b T/m )^{-1/2}$, and $\lambda$ denotes the mean-free-path based on the number density of the left side of the initial discontinuity
\begin{flalign}
\lambda = \frac{1}{\sqrt{2} n_L \pi \sigma^2}~.
\end{flalign}
\noindent  Properties of Argon with molecular mass $m=6.633521\times 10^{-26}\ \mathrm{kg}$ and Hard-Sphere diameter of $\sigma=3.405 \times 10^{-10} \ \mathrm{m}$ as an ideal monatomic gas are assumed. The viscosity follows $\mu=5 \sqrt{mk_bT/\pi}/(16 \sigma^2)$ and the heat conductivity $\kappa= 15 k_b \mu/(4 m)$. In all test cases, the time step size is  fixed to the value of $\Delta t = 2.01\times 10^{-9}\ {\color{ms}\mathrm{s}}$. The SPH particles adopt identical weight. Moreover, SPH and hybrid solution algorithms are initialized with {\color{ms_rev}$80,000$} particles in the left side of the domain and corresponding number to the right side. DSMC simulations are performed using the statistical weights $F_N \in\{ 10^{13}, 10^{12}, 10^{11} \} \times {\color{ms_rev}4.71092}$ which {\color{ms_rev}corresponds} to $\rho_0 \in \{10^{-4}, 10^{-5}, 10^{-6} \}\ {\textrm{kg.m}^{-3}}$.
\\ \ \\
Three cases of the initial densities with the same termination time, provides different levels of non-equilibrium. First, the solution for the case with $\rho_{L}=10^{-4}\ \mathrm{kg.m^{-3}}$  obtained from full DSMC, full SPH, and hybrid DSMC-SPH based on GP/ANN MED estimates are presented in Fig.~\ref{fig:shock_tube_GPR_1e-4}-\ref{fig:shock_tube_NN_1e-4}. For comparisons, the evolution of the Fisher information distance from equilibrium, along with the cell flags at four time intervals are shown in Fig.~\ref{fig:shock_tube_Ifish_GPR_1e-4}. Although DSMC solution is subject to noise, overall good agreement between the hybrid solution algorithm and full DSMC results is observed; considering both GP and ANN estimates of MED.

\begin{figure}[t]
\centering
	\includegraphics[scale=0.9]{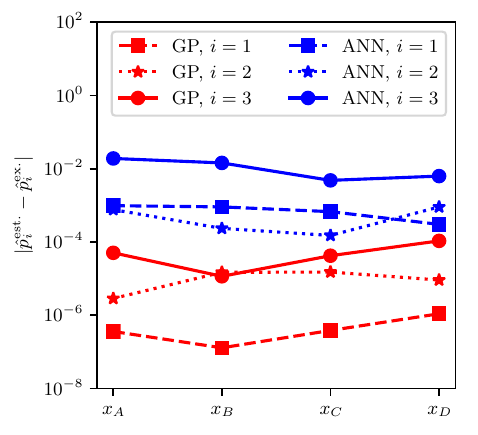}
	\caption{Absolute value error of  moments of the estimated $f^\lambda_3$ using GP and ANN at four locations $x_A=0.45\ \mathrm{m}$, $x_B=0.5\ \mathrm{m}$, $x_C=0.55 \ \mathrm{m}$, and $x_D=0.6\ \mathrm{m}$  for the Sod's shock tube hybrid simulation at $t = 2.01\times 10^{-4}\ \mathrm{s}$.}
  \label{fig:error_shock_mom}
\end{figure}

\begin{figure}
  \centering
  \begin{minipage}[l]{0.45\textwidth}
\centering
    \includegraphics[scale=0.8]{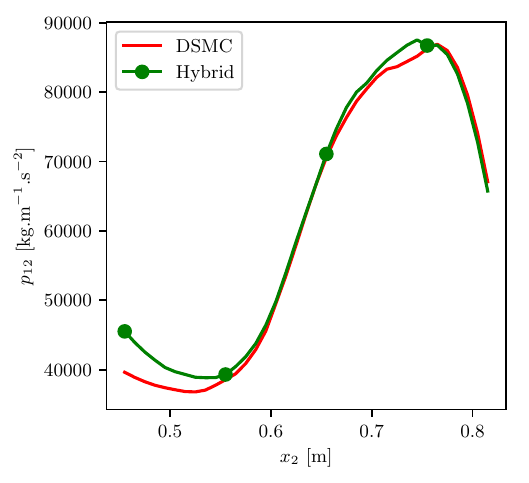}
    \subcaption{}
\end{minipage}
\begin{minipage}[c]{0.45\textwidth}
\centering
    \includegraphics[scale=0.8]{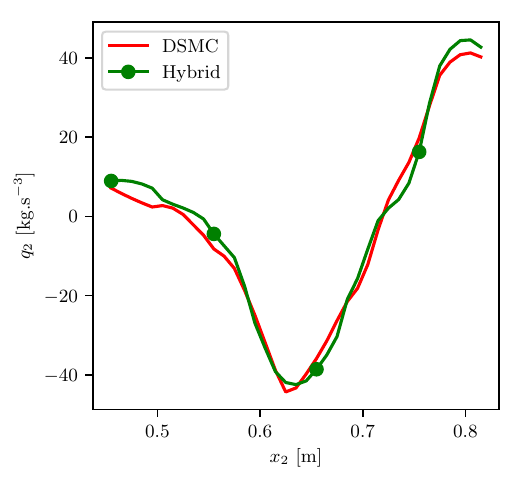}
    \subcaption{}
\end{minipage}
  \caption{\color{ms_rev} The kinetic pressure $p_{12}$ and the heat flux $q_2$ obtained from hybrid and DSMC solution algorithms at time $t = 8.39\times 10^{-4}\ \mathrm{s}$.}
  \label{fig:shock_p12q2}
\end{figure}

\begin{figure}
  \centering
    \begin{minipage}[l]{0.45\textwidth}
\centering
    \includegraphics[scale=0.9]{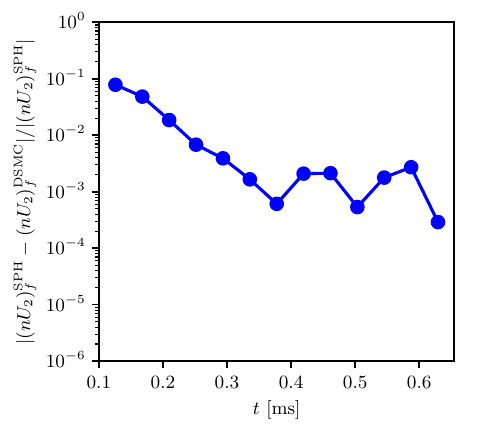}
    \subcaption{}
\end{minipage}
\begin{minipage}[l]{0.45\textwidth}
\centering
    \includegraphics[scale=0.9]{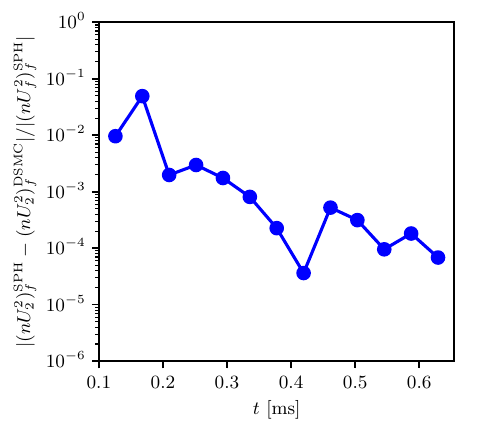}
        \subcaption{}
\end{minipage}
  \begin{minipage}[l]{0.45\textwidth}
\centering
    \includegraphics[scale=0.9]{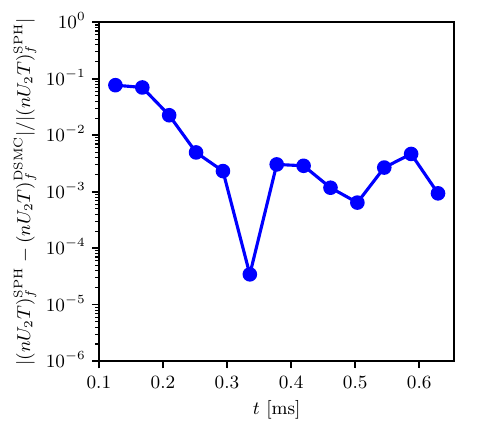}
        \subcaption{}
\end{minipage}
\begin{minipage}[l]{0.45\textwidth}
\centering
        \includegraphics[scale=0.9]{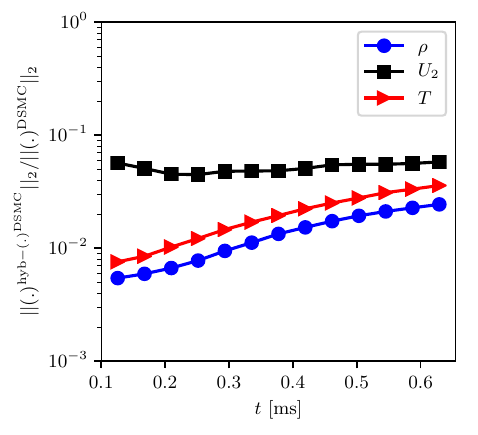}
            \subcaption{}
\end{minipage}
\caption{{\color{ms_rev} Shock tube with $\rho_0=10^{-4}\ \mathrm{kg.m^{-3}}$. Evolution of the relative error in numerical approximation of fluxes $nU_2$, $nU_2^2$ and $nU_2T$ shown in (a), (b) and (c), respectively, are obtained from DSMC and SPH side of the hybrid solution algorithm. Furthermore, the evolution of relative $L^2$-norm error of conserved quantities in comparison with DSMC solution is plotted in (d).}}
\label{fig:err_flux_L2}
\end{figure}

\begin{figure}
  \centering
\includegraphics[scale=0.9]{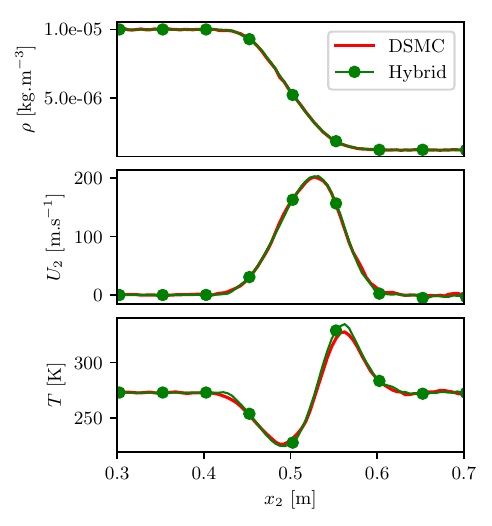}
\includegraphics[scale=0.9]{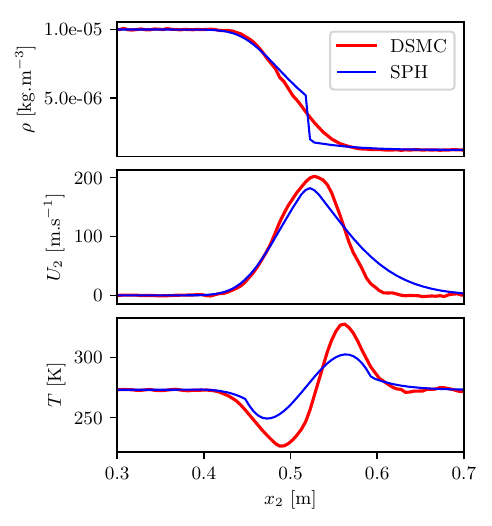}
  \caption{Sod's shock tube with initial values of $(\rho_L, U_{2,L}, T_L) = (10^{-5} ~\mathrm{kg.m^{-3}}, 0 ~\mathrm{m.s^{-1}}, 273 ~\mathrm{K})$ and $(\rho_R, U_{2,R}, T_R) = (0.125\times 10^{-5} ~\mathrm{kg.m^{-3}}, 0~\mathrm{m.s^{-1}}, 273~\mathrm{K})$ at $t = 2.01\times 10^{-4}\ \mathrm{s}$. The profiles of the density $\rho$, the mean velocity $U_2$, and the temperature $T$ are obtained from SPH, DSMC, and the devised hybrid DSMC-SPH equipped with {\color{ms_rev}GP} estimator of MED.}
  \label{fig:shock_tube_GPR_1e-5}
\end{figure}

\begin{figure}
  \centering
\includegraphics[scale=0.9]{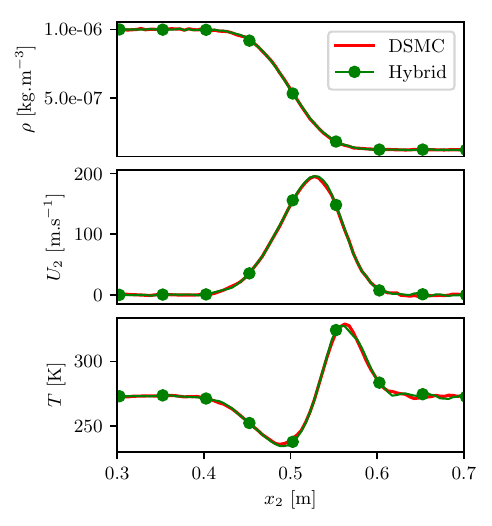}
\includegraphics[scale=0.9]{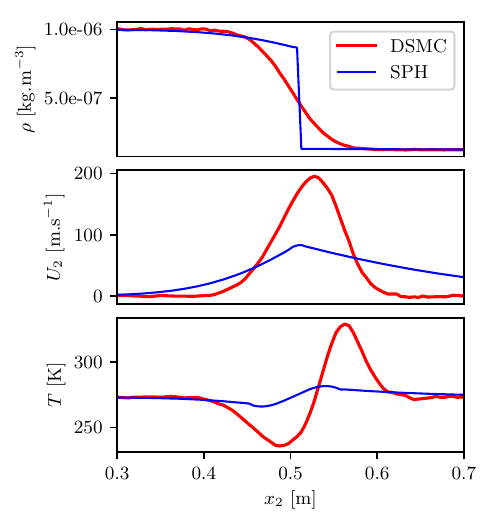}
  \caption{Sod's shock tube with initial values of $(\rho_L, U_{2,L}, T_L) = (10^{-6} ~\mathrm{kg.m^{-3}}, 0 ~\mathrm{m.s^{-1}}, 273 ~\mathrm{K})$ and $(\rho_R, U_{2,R}, T_R) = (0.125\times 10^{-6} ~\mathrm{kg.m^{-3}}, 0~\mathrm{m.s^{-1}}, 273~\mathrm{K})$ at $t = 2.01\times 10^{-4}\ \mathrm{s}$. The profiles of the density $\rho$, the mean velocity $U_2$, and the temperature $T$ are obtained from SPH, DSMC, and the devised hybrid DSMC-SPH equipped with {\color{ms_rev}GP} estimator of MED.}
  \label{fig:shock_tube_GPR_1e-6}
\end{figure}

\noindent To gain a better insight into the accuracy of data-driven GP/ANN estimators of MED, we extract moments from four probe positions $x_A=0.45\ \mathrm{m}$, $x_B=0.5\ \mathrm{m}$, $x_C=0.55 \ \mathrm{m}$, and $x_D=0.6\ \mathrm{m}$ at the terminal time. Instead of computing the exact MED solution,  we have computed the moments of the estimated MEDs and compared them with respect to the input moment. As it can {\color{ms_rev} be} seen in Fig.~\ref{fig:error_shock_mom}, the predictions based on the GP regression shows a better accuracy in comparison to the one from ANN. Consequently, for the remaining test cases, only the trained GP method is employed in the hybrid solution algorithm. \\ \ \\
{\color{ms_rev}
Furthermore, higher order moments of our hybrid solution algorithm equipped with the GP estimator of MED is compared with DSMC results. As shown in Fig.~\ref{fig:shock_p12q2}, the pressure $p_{12}$ and the heat flux $q_2$ obtained from the hybrid algorithm are in good agreement with those obtained from DSMC. Also, in order to investigate the conservation laws across the interface,  the relative error in estimating the advective fluxes at the interface from DSMC and SPH sides, as well as the overall relative $L^2$-norm error of conserved quantities are depicted in Fig.~\ref{fig:err_flux_L2}, where again reasonable accuracy (subject to statistical noise) is visible. 
\\ \ \\}
\noindent Next, more rarefied scenarios are considered for the Sod's shock tube with $\rho_0=10^{-5}\ \mathrm{kg.m^{-3}}$ and $\rho_0=10^{-6} \ \mathrm{kg.m^{-3}}$, as the results are shown in Figs.~\ref{fig:shock_tube_GPR_1e-5} and \ref{fig:shock_tube_GPR_1e-6}, respectively. For lower density while keeping ratio $\rho_L/\rho_R$ constant, the solution of SPH departs further from DSMC results. However, the hybrid solution shows a good accuracy in comparison with the benchmark DSMC result.
\subsection{Performance of hybrid solution algorithm}

\begin{figure}[t]
\centering
\begin{subfigure}{0.4\columnwidth}
	\includegraphics[scale=0.8]{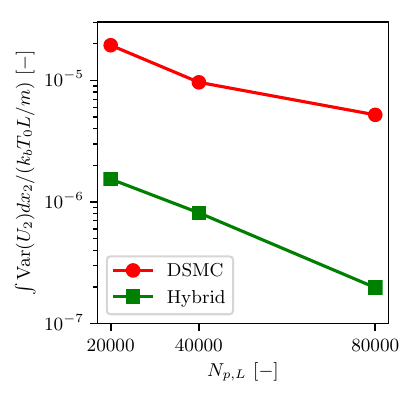}
	\caption{Spatial averaged noise of $U_2$ estimations using DSMC and devised hybrid algorithm for the case of $\rho_0 = 10^{-4}\ \mathrm{kg.m^{-3}}$ at $t = 1 \times 10^{-4}\ \mathrm{s}$, where $10$ independent simulation results were employed to obtain the variance at each point in space. Here $N_{p,L}$ indicates the number of SPH particles initialized at the left side of the discontinuity for the hybrid method, while for the full DSMC simulation 200 times more particles i.e. $N_{p,L}^{\mathrm{DSMC}}=200 N_{p,L}$ were generated at the left portion of the domain.}
\end{subfigure}
\ 
\begin{subfigure}{0.4\columnwidth}
	\includegraphics[scale=0.9]{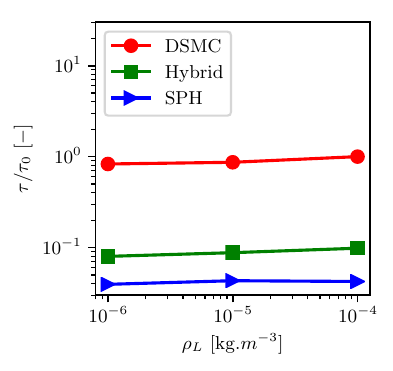}
	\caption{Normalized CPU time $\tau$ for the Sod's shock tube problem using DSMC, SPH and devised hybrid DSMC-SPH with $t_\text{final} = 1\times 10^{-4}\ \mathrm{s}$. Here $N_{p,L}=40,000$ particles were initialized on the left side of the discontinuity for the SPH and hybrid method, while $N_{p,L}^{\mathrm{DSMC}}=200 N_{p,L}$ used initially for full DSMC simulation. The execution time $\tau$ is normalized by $\tau_0$ indicating DSMC execution time of the case $\rho_0 = 10^{-4}\ \mathrm{kg.m^{-3}}$.}
\end{subfigure}
  \caption{Noise and cost study of DSMC, SPH and DSMC-SPH solution algorithms for the Sod's shock tube test case}
  \label{fig:noise_cost}
\end{figure}

\noindent As more cells remain in the continuum limit and assigned to the SPH solver in the hybrid approach, much less noise compared to the full DSMC simulation is expected. Moreover, even though the hybrid solution algorithm introduces an overhead mainly due to particle samplings from MED estimates, still good computational efficiency can be obtained since typically only a small fraction of cells switch to DSMC at a given time step. Computational speedups along with noise reductions associated with the hybrid results are reported in Fig.~\ref{fig:noise_cost}.

\section{Statistical errors}
\begin{figure}
  \centering
\includegraphics[scale=0.9]{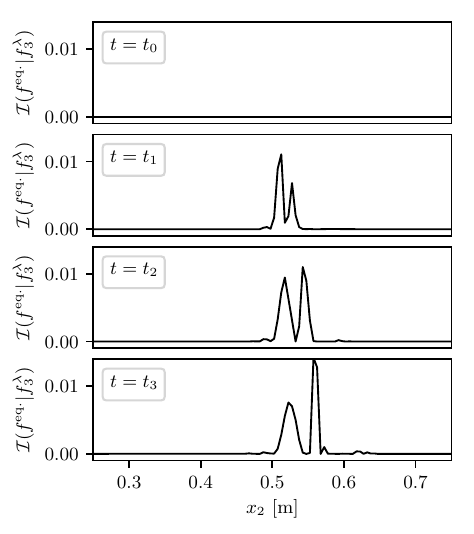}
\includegraphics[scale=0.9]{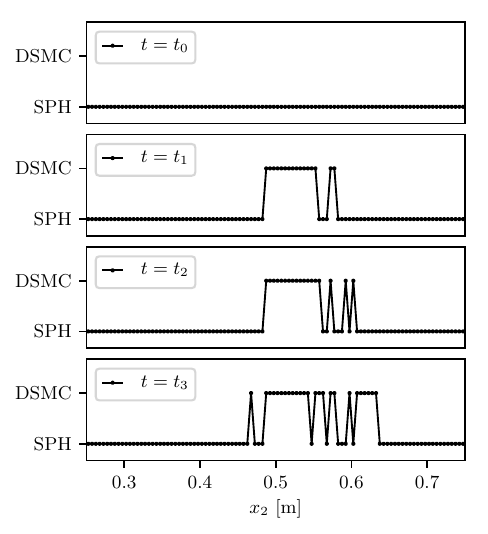}
  \caption{\color{ms_rev1}Evolution of the Fisher information distance $\mathcal{I}(f^{\mathrm{eq.}}|f_3^\lambda)$ at $t \in \{ 0,0.419, 0.839, 1.25 \}\times 10^{-4}\ \mathrm{s}$ and the assigned DSMC/SPH solver in the hybrid method for the Sod's shock tube test case with $\rho_0=10^{-4}\ \mathrm{kg.m^{-3}}$, shown at left and right, respectively. Here, GP estimate of MED with breakdown parameter $\epsilon=8\times10^{-5}$ is deployed.}
  \label{fig:shock_tube_Ifish_GPR_1e-4_noise}
\end{figure}

\begin{figure}
  \centering
  \includegraphics[scale=0.9]{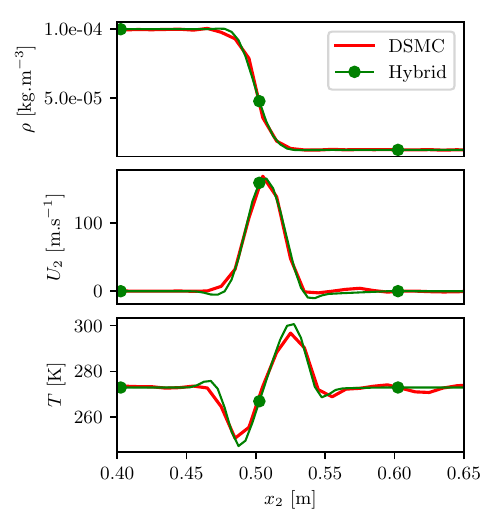}
\includegraphics[scale=0.9]{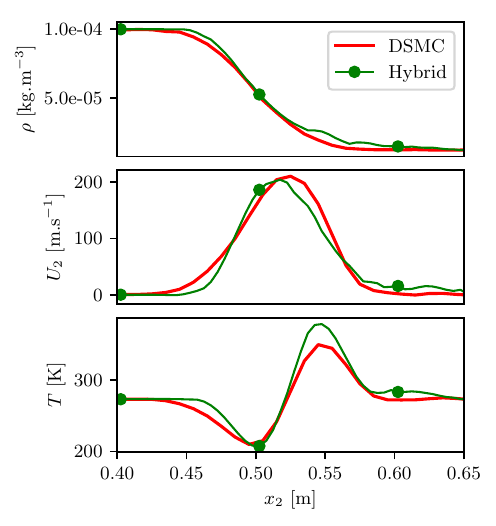}
  \caption{\color{ms_rev1}Sod's shock tube with initial values of $(\rho_L, U_{2,L}, T_L) = (10^{-4} ~\mathrm{kg.m^{-3}}, 0 ~\mathrm{m.s^{-1}}, 273 ~\mathrm{K})$ and $(\rho_R, U_{2,R}, T_R) = (0.125\times 10^{-4} ~\mathrm{kg.m^{-3}}, 0~\mathrm{m.s^{-1}}, 273~\mathrm{K})$ at $t \in\{ 0.419,\ 1.25\}\times 10^{-4}\ \mathrm{s}$ shown in left and right, respectively. The profiles of the density $\rho$, the mean velocity $U_2$, and the temperature $T$ are obtained from  DSMC and the devised hybrid DSMC-SPH equipped with GP estimator of MED. Here, a larger breakdown parameter $\epsilon=8\times10^{-5}$ is chosen to reach more frequent information exchange between kinetic and continuum solution.}
  \label{fig:shock_tube_noise}
\end{figure}

\noindent The data-driven MED   closure for the moment problem resolves the closure issue with continuum-to-kinetic transformation as well as estimation of the continuum breakdown, yet the Monte Carlo noise is inherited in the kinetic-to-continuum back transformation. To highlight this limitation, here we look at the simulation results of the sod-shock tube explained in \S~\ref{sec:shock_tube} for $\rho_0=10^{-4}\ \mathrm{kg.m}^{-3}$ with a higher tolerance for the continuum breakdown estimation $\epsilon=8\times10^{-5}$, in order to observe more exchange between kinetic and continuum solver, as it can be seen in Fig.~\ref{fig:shock_tube_Ifish_GPR_1e-4_noise}. More frequent information exchange between the solvers leads to further error in the overall solution. The noise introduced from the Monte Carlo method to the SPH,  leads to accumulation of error in the hybrid approach, as can be seen from the evolution of  density, temperature and velocity profiles in Fig.~\ref{fig:shock_tube_noise}. We anticipate that integrating a noise reduction scheme with the presented hybrid approach improves the accuracy of the scheme.

\section{Conclusion and outlook}
\label{sec:conc_disc_outlook}
\noindent This study explores some potential of data-driven methodologies to accelerate multi-scale computations, relevant to hybrid kinetic-continuum settings. Two main challenges of interface treatment and switching criterion, associated with coupling of continuum and kinetic solvers are addressed. It is shown that a practical and accurate probability density estimator can circumvent these challenges and thus make efficient hybrid algorithms more accessible. By leveraging GP and ANN regression schemes, efficient density estimations based on the Maximum-Entropy condition are derived. Accordingly, a hybrid solution algorithm based on fast and accurate MED estimators is devised and tested in one-dimensional shock tube scenarios. Based on detailed comparisons, the GP regression model was picked as the method of choice for the studied scenario. Very good agreement with respect to the DSMC benchmark along with significant speed-up are observed. While the setting considered here was one-dimensional, more dimensions can be included in a straightforward manner, given that all presented derivations are independent of the number of dimensions. Furthermore, even though only DSMC-SPH coupling was addressed throughout this work, the proposed approach motivates the use of data-driven MED for coupling higher-order moment systems coupled with kinetic solution algorithms.

\section*{Acknowledgments}
\noindent Hossein Gorji acknowledges the funding provided by the Swiss National Science Foundation under the grant number 174060. Furthermore, Mohsen Sadr acknowledges the scholarship provided by the German Academic Exchange Service (DAAD) identifiable with number 57438025-91749111.

\bibliographystyle{unsrt}
\bibliography{refs}

\end{document}